\documentstyle[graphics]{cupconf}

\ifoldfss
\else
  \ifnfssone
    \newmathalphabet{\mathit}
      \addtoversion{normal}{\mathit}{cmr}{m}{it}
      \addtoversion{bold}{\mathit}{cmr}{bx}{it}
    \newmathalphabet{\mathcal}
      \addtoversion{normal}{\mathcal}{\cmsy}{m}{n}
    \else
    \ifnfsstwo
    \fi
  \fi
\fi

\def\ie{\mbox{i.e.\ }}
\def\gtrsim{\mathrel{\hbox{\rlap{\hbox{\lower4pt\hbox{$\sim$}}}\hbox{$>$}}}}

\def\lesssim{\mathrel{\hbox{\rlap{\hbox{\lower4pt\hbox{$\sim$}}}\hbox{$<$}}}}

\def\hexnumber#1{\ifcase#1 0\or1\or2\or3\or4\or5\or6\or7\or8\or9\or
 A\or B\or C\or D\or E\or F\fi }

\makeatletter
\ifx\CUP@mtlplain@loaded\undefined
\else
\fi
\makeatother
 \makeatletter
 \ifx\CUP@mtlplain@loaded\undefined
   \font\tenbmi=cmmib10 at 10pt
   \font\sevenbmi=cmmib10 at 7pt
   \font\fivebmi=cmmib10 at 5pt

   \newfam\bmifam
   \textfont\bmifam=\tenbmi
   \scriptfont\bmifam=\sevenbmi
   \scriptscriptfont\bmifam=\fivebmi
   
 \fi
 \makeatother
\ifnfsstwo

\fi
\ifnfssone

\fi
\ifoldfss

\fi

\mathchardef\varLambda="0103

\makeatletter
\ifx\CUP@mtlplain@loaded\undefined
\else
\fi
\makeatother
\makeatletter
\ifx\CUP@mtlplain@loaded\undefined
  \font\tenbms=cmbsy10
  \font\sevenbms=cmbsy10 at 7pt
  \font\fivebms=cmbsy10 at 5pt
  \newfam\bmsfam
  \textfont\bmsfam=\tenbms
  \scriptfont\bmsfam=\sevenbms
  \scriptscriptfont\bmsfam=\fivebms

  \edef\bsy@{\hexnumber\bmsfam}
  \mathchardef\bnabla="0\bsy@72
\fi
\makeatother
\def\eg{{e.g.\ }}

\def\etal{\mbox{\it et al.}}

\newcommand{\cmcub}{~cm$^{-3}$}

\newcommand{\Ms}{~M$_{\odot}$}
\newcommand{\Ls}{~L$_{\odot}$}

\newcommand{\qh}{$Q({\rm{H^{0}}})$}

\newcommand{\Te}{$T_{e}$}
\newcommand{\Tstar}{$T_{\star}$}

\newcommand{\Mstar}{$M_{\star}$}

\newcommand{\Mneb}{$M_{\rm{neb}}$}

\newcommand{\Ha}{H$\alpha$}
\newcommand{\Hb}{\ifmmode {\rm H}\beta \else H$\beta$\fi}

\newcommand{\hii}{H~{\sc ii}} 
\newcommand{\hei}{He~{\sc i}} 
\newcommand{\Hei}{He~{\sc i} $\lambda$5876} 
\newcommand{\heii}{He~{\sc ii}} 
\newcommand{\Heii}{He~{\sc ii} $\lambda$4686} 
\newcommand{\Ciii}{C~{\sc iii}] $\lambda$1909}
 
\newcommand{\Nii}{[N~{\sc ii}] $\lambda$6584} 
\newcommand{\nii}{[N~{\sc ii}]} 
\newcommand{\Oi}{[O~{\sc i}] $\lambda$6300} 
 
\newcommand{\Oii}{[O~{\sc ii}] $\lambda$3727} 
\newcommand{\oii}{[O~{\sc ii}]} 
\newcommand{\Oiii}{[O~{\sc iii}] $\lambda$5007} 
\newcommand{\Oiiit}{[O~{\sc iii}] $\lambda$4363} 
\newcommand{\oiii}{[O~{\sc iii}]}

\newcommand{\nev}{[Ne~{\sc v}]}
\newcommand{\Sii}{[S~{\sc ii}] $\lambda$6716, $\lambda$6731} 
 
\newcommand{\Siii}{[S~{\sc iii}] $\lambda$9532}

\newcommand{\Ariv}{[Ar~{\sc iv}] $\lambda$4711+4740}

\newcommand{\Oiiitonea}{[O\,{\sc{iii}}]\,$\lambda$52\,$\mu$m}
\newcommand{\Oiiitoneb}{[O\,{\sc{iii}}]\,$\lambda$88\,$\mu$m}

\newcommand{\rOii}{[O~{\sc ii}] $\lambda$3726/3729}
\newcommand{\rOiii}{[O~{\sc iii}] $\lambda$4363/5007}
\newcommand{\rNii}{[N~{\sc ii}] $\lambda$5755/6584}
\newcommand{\rSiii}{[S~{\sc iii}] $\lambda$6312/9532} 
\newcommand{\rSii}{[S~{\sc ii}] $\lambda$6731/6717} 
\newcommand{\rAriv}{[Ar~{\sc iv}] $\lambda$4740/4711}

\newcommand{\Ho}{H$^{0}$}
\newcommand{\Hp}{H$^{+}$}
\newcommand{\He}{He$^{0}$}
\newcommand{\Hep}{He$^{+}$}
\newcommand{\Hepp}{He$^{++}$}
\newcommand{\Cp}{C$^{+}$}
\newcommand{\Cpp}{C$^{++}$}

\newcommand{\Np}{N$^{+}$}
\newcommand{\Npp}{N$^{++}$}

\newcommand{\Op}{O$^{+}$}
\newcommand{\Opp}{O$^{++}$}
\newcommand{\Oppp}{O$^{+++}$}
\newcommand{\Nep}{Ne$^{+}$}
\newcommand{\Nepp}{Ne$^{++}$}

\newcommand{\Spp}{S$^{++}$}
\newcommand{\Arpp}{Ar$^{++}$}

\title[Abundances in \hii\ regions and planetary nebulae]{Abundance 
determinations in \hii\ regions and planetary nebulae}

\author[G. Stasi\'{n}ska]%
{G\ls R\ls A\ls \.{Z}\ls Y\ls N\ls A\ns S\ls T\ls A\ls S\ls I\ls \'{N}\ls 
S\ls K\ls A\ls
}

\affiliation{Observatoire de Paris-Meudon, 5, place Jules Janssen, 
92195 
Meudon cedex, France}

\begin{document}
\ifnfssone
\else
  \ifnfsstwo
  \else
    \ifoldfss
      \let\mathcal\cal
      \let\mathrm\rm
      \let\mathsf\sf
    \fi
  \fi
\fi

\maketitle

\begin{abstract}
The methods of abundance determinations in \hii\ regions and 
planetary nebulae are described,  with emphasis on 
the underlying assumptions and inherent problems. Recent results on 
abundances in Galactic \hii\ regions and in Galactic and extragalactic 
Planetary Nebulae are 
reviewed.

\end{abstract}

\hii\ regions are ionized clouds of gas associated with zones of recent star 
formation.  They are powered by one, a few, or a cluster of massive 
stars (depending on the resolution at which one is working).  The 
effective temperatures \Tstar\ of the ionizing stars lie in the range 
35\,000 -- 50\,000~K.  The nebular geometries result from the 
structure of the parent molecular cloud.  Stellar winds, at evolved 
stages, may produce ring-like structures, but the morphology of \hii\ 
regions is generally rather complex on all scales.  Typical hydrogen 
densities $n$ are $10^{3}$ -- $10^{4}$\cmcub\ for compact \hii\ 
regions.  The average densities in giant extragalactic \hii\ regions 
 are lower, typically $10^{2}$\cmcub\, since giant  \hii\ regions encompass also 
zones of diffuse material.  The total supply of nebular gas is 
generally large, so that all (or at least a significant fraction) of 
the ionizing photons are absorbed.

Planetary nebulae (PNe) are evolutionary products of so-called intermediate 
mass stars (initial masses of 1 -- 8\Ms) as they progress from the 
asymptotic giant branch (AGB) to the white dwarf stage.  It is the 
interaction of the slow AGB wind with the fast post-AGB wind which 
produces the nebula.  Because the ionizing star is also the remnant of 
the PN progenitor, the morphology is much simpler that in the case of 
\hii\ regions, although not all PNe are round!  The temperature of the 
central star -- or nucleus -- can be much higher than that of main 
sequence massive stars, reaching values of the order 200\,000~K for a 
remnant of about 0.6\Ms.  The densities of the brightest (and 
therefore best studied) PNe are around $10^{3}$ -- $10^{5}$\cmcub.  
PNe of lower densities, corresponding to more evolved stages, are 
fainter and therefore less observed.  The amount of nebular gas is not 
always sufficient to trap all the stellar ionizing photons, 
and a significant part of these may leak out from the nebula.
 
This brief introduction points at two things.  One is that the ionized 
plasmas in \hii\ regions and PNe are similar from the physical point of view, 
and therefore can be analyzed with similar techniques (although the 
range of physical conditions is somewhat different).  The other is that the astrophysical significance of the 
chemical composition in these two classes of objects is not the same. \hii\ regions probe the state of the gas at the birth of 
massive stars (\ie a few Myr ago).  The status of the chemical 
composition of PN envelopes is more complex.  Some constituents have 
not been changed and reflect the state of the gas out of which the 
progenitor of the PN was formed, 10$^{8}$~yr ago or more.  Other 
elements, such as carbon and nitrogen, have had their abundances 
strongly affected by nucleosynthesis and mixing processes in the 
progenitor, and therefore probe the evolution of intermediate mass 
stars.

The text presented below is based on lectures given at the XIII 
Canary Islands Winterschool on Cosmochemistry, where I have been asked 
to review the status of abundances in planetary nebulae (both 
Galactic and extragalactic) and in 
Galactic \hii\ regions.  Abundances in extragalactic \hii\ regions 
were treated by Don Garnett, and the determination of the primordial 
helium abundance using low metallicity \hii\ galaxies was discussed by 
Gary Steigman. In my lectures, I have emphasized the methods for abundance 
determinations in 
ionized nebulae.  In this respect, giant extragalactic \hii\ regions 
provide interesting complementary information and methods used for 
giant \hii\ regions were included for completeness.

The scope of this article is as follows. Section 1 summarizes the basic 
physics of photoionized nebulae, Section 2 presents the different 
families 
of methods for abundance determinations. Section 3 discusses the 
various sources of uncertainties. Section 4 outines some important 
recent results on abundances in the Milky Way \hii\ regions, including ring 
nebulae. Section 5 presents a selection of recent results on abundances in planetary 
nebulae, that are relevant to our understanding of the chemical history of galaxies 
or of the nucleosynthesis in intermediate mass stars. Due to limited 
space (and limited knowledge!), Sections 4 and 5 are not to be 
taken for extensive reviews. A large amount of interesting work 
could not be mentioned here. This text is rather to be understood as a 
guide line for the astronomer interested in nebular abundances, either 
to embark on his own abundance determinations or to be 
able to better understand the literature on this topic. The papers 
quoted below were preferably chosen among recent studies
 published in refereed journals. A few pioneering, older studies 
 are occasionally mentioned.

\section {Basic physics of photoionized nebulae}

Excellent introductions are provided in textbooks such as those of Spitzer 
(1978), Aller (1984) or Osterbrock (1989).  Here, we simply emphasize the 
properties to bear in mind when dealing with abundance determinations.

 \subsection {Ionization and recombination}
 
  \subsubsection {Global ionization budget}
 
Consider a source of photons surrounded by a cloud of nebular gas.  
The gas particles are ionized by those photons with energies above the 
ionization threshold.  Once ionized, the particles tend to recombine 
with the free electrons, and an equilibrium stage is eventually 
established in which the rate of ionization equals the rate of 
recombination for each species.

Closer to the source, the density of ionizing photons is larger, 
therefore the resulting ionization state of the gas is higher.  If 
there is enough nebular matter, all the ionizing photons can be absorbed, 
producing an ionization bounded nebula.  If not, the nebula is called 
density bounded.

It is the most abundant species, (H and He in general, but this could be 
C, N, O in hydrogen-poor material) which absorb most of the Lyman continuum 
photons from the ionizing source, and thus define the size of the 
ionized region in the ionization bounded case.

In an ionization bounded nebula purely composed of hydrogen, the total 
number of recombinations per unit time balances the total 
number of photons with energies above 13.6~eV emitted  per unit time either 
 by the star, or during recombination to the ground level. One has: 

 \begin{equation}
Q({\rm{H^{0}}}) + \int_{}^{} n({\rm{H^{+}}}) n_{e} \epsilon \alpha_{1}({\rm{H}},T_{e})
dV = \int_{}^{} n({\rm{H^{+}}}) n_{e} \epsilon \alpha_{tot}({\rm{H}}, 
T_{e}) {\rm d}V, 
\end{equation}
where \qh\ is the total number of photons with energies above 
13.6~eV emitted by the star per second; $n({\rm{H^{+}}})$ is the number 
 density of H ions, $n_{e}$ is the electron density, $\epsilon$ is the 
 volume filling factor of the nebular gas; $\alpha_{1}({\rm{H}}, T_{e})$ is 
 the H recombination coefficient to the ground level 
 while  $\alpha_{tot}({\rm{H}},T_{e})$ is the total H recombination 
 coefficient, which are both roughly inversely proportional to the electron 
 temperature \Te.
The integrations are 
 performed over the nebular volume.

In the case of a constant density nebula with constant filling factor, 
the radius of the ionized region, or Str\"omgren radius is then:
\begin{equation}
R_{S} = \left(\frac{3 Q({\rm{H^{0}}})}{4 \pi \epsilon n_{e}^{2} \alpha_{B}({\rm{H}}, 
T_{e})}\right)^{1/3},
\end{equation}
where $\alpha_{B}({\rm{H}},T_{e})$ is the H recombination coefficient to the excited 
 states (in this equation, $T_{e}$ represents an average electron temperature of the 
nebula).
 The thickness of the transition region between the fully ionized zone
and the neutral zone is approximately one mean free path of an 
ionizing photon $d = 1/ n({\rm{H^{0}}}) \alpha_{\nu}$, where 
$\alpha_{\nu}$ is the hydrogen photoionization cross section at the 
typical frequency of the photons reaching the ionization front.  This 
thickness is generally much smaller than the size of the nebula and 
justifies the concept of a Str\"{o}mgren sphere.  There are however 
cases when the transition region might be extended, such as in 
diffuse media or when the ionizing radiation field contains a large 
amount of X-ray photons (which are less efficiently absorbed by 
hydrogen).

During the recombination process captures to the excited levels 
decay to lower levels by radiative transitions.  The total luminosity of 
the \Hb\ line is thus
\begin{equation}
L_{\rm{H}\beta} = \int_{}^{} n({\rm{H^{+}}}) n_{\rm{e}} \epsilon 
4 \pi j_{\rm{H}\beta}(T_{e}){\rm d}V,
\end{equation}
where $j_{\rm{H}\beta}(T_{e})$  is the emission 
coefficient of \Hb\ and is roughly proportional to 
$\alpha_{B}({\rm{H}})$. Therefore the total 
luminosity in \Hb\ in an ionization bounded nebula is 
a direct measure of \qh. At \Te\ = $10^{4}$~K, it is given by: 
\begin{equation}
L_{\rm{H}\beta} = 4.8~10^{-13} Q({\rm{H^{0}}}) 
{\rm erg~s^{-1}}.
\end{equation}

In the case of a density bounded nebula, though, some ionizing photons 
escape and $L_{(\rm{H}\beta)}$ is then given by:
\begin{equation}
L_{\rm{H}\beta} = 1.5~10^{32} (T_{e}/ 10^{4}) ^{-0.9} M_{neb} 
<n> {\rm erg~s^{-1}},
\end{equation}
 where \Mneb\ is the nebular mass in solar units and $<n>$ 
 is defined as:
\begin{equation}
<n> = \int_{}^{} n^{2}  \epsilon {\rm d}V / \int_{}^{} n\epsilon {\rm d}V,
\end{equation}
 assuming that in the nebula $ n(\rm{H^{+}})$ = $n_{e}$ = 
 $n(\rm{H})$ $\equiv$ $n$.

Thus, in the density bounded case, the total \Hb\ luminosity does not 
say anything about \qh, except that \qh\ has to be larger than the value 
required to obtain the observed luminosity in \Hb.  For a given total nebular 
mass, $L_{\rm{H}\beta}$ is larger for denser nebulae, since recombinations 
are then more frequent.

  For nebulae composed of pure hydrogen, the maximum ionizable mass of 
  gas for a given value of \qh\ is, 
  at \Te\ = $10^{4}$~K:  
\begin{equation}
M_{ion} = 3.2 10^{-45} Q({\rm{H^{0}}}) / <n> \rm{M}_{\odot}.
\end{equation}

The following table gives the values of $M_{ion}$ for a 
typical PN, an \hii\ region ionized by an O7 star, and a giant \hii\ 
region ionized by a cluster of stars representing a total mass of 
$10^{4}$~\Ms\ (a Salpeter mass function 
is assumed and the star masses range between 1 and 100~\Ms).

\begin{table*}
\begin{flushleft}
\begin{tabular}{lllll}
\hline               
    & \qh\
    & \Mstar\ 
     & $M_{\rm{ion}}$  & $M_{\rm{ion}}$  \\     
     \hline               
    &  
    &  
     & $n$ = $10^{2}$~\cmcub\  & $n$ = $10^{4}$~\cmcub\ \\
\hline 
planetary nebula & 3\,10$^{47}$ ph~s$^{-1}$ & 0.6\Ms  & 10\Ms  & 
10$^{-1}$\Ms \\
single star H~{\sc ii} region & 3\,10$^{48}$ ph~s$^{-1}$ & 30\Ms & 10$^{2}$\Ms & 1\Ms \\
giant H~{\sc ii} region & 3\,10$^{50}$ ph~s$^{-1}$ & 10$^{4}$\Ms & 
10$^{4}$\Ms  & 10$^{2}$\Ms \\
\hline                               
\end{tabular}
\end{flushleft}
\label{Table1}
\caption{Typical masses of the ionizing stars (or star clusters) and 
maximum nebular ionizable masses}
\end{table*}

The surface brightness of an object is an important parameter from the 
observational point of view.  Indeed, for extended objects, it 
determines the detectability or the quality of the spectra.  For 
illustrative purposes, let us consider here the simple case of an 
homogeneous sphere and define:
\begin{equation}
S_{\rm{H}\beta} = F_{\rm{H}\beta} / (\pi \theta^{2}) = L_{\rm{H}\beta} 
/ (4 \pi^{2} R_{neb}^{2}),
\end{equation}
where $F_{\rm{H}\beta}$ is the observed \Hb\ flux, $\theta$ is the angular 
radius of the nebula and $R_{neb}$ its physical radius.  
With the help of the previous equations one obtains for the ionization 
bounded case:
\begin{equation}
S_{\rm{H}\beta} \propto (Q({\rm H^{0}}) n^{4} \epsilon^{2})^{1/3}
\end{equation}
and for the density bounded case: 
\begin{equation}
S_{\rm{H}\beta} \propto (M_{\rm{neb}} n^{5} \epsilon^{2})^{1/3}.
\end{equation}
Thus better data will be obtainable for objects of higher densities, and 
objects with higher $M_{\rm{neb}}$ or \qh.

The number fractions of He and heavy elements (C, N, O\ldots)\footnote{It is 
a tradition in nebular studies, to refer to 
elements other than H and He as ``heavy elements'' or ``metals''.} in real 
nebulae are about 10\% and 0.1\% respectively.  
Helium, although ten times less abundant than hydrogen, is the 
dominant source of absorption of photons at energies above 24.4~eV. 
For order of magnitudes estimates, however, the formulae given above 
can still be used, since each recombination of He roughly  produces one 
photon that can subsequently be absorbed only by hydrogen.  The same 
remark generally holds for photons above 54.4~eV in the spectra of 
PNe with hot nuclei (see however Stasi\'{n}ska \& 
Tylenda 1986).  Naturally, for detailed studies, 
a photoionization modelling is necessary that 
takes into account properly the transfer of the photons arising from 
the recombination to \He\ and \Hep.

  \subsubsection{The ionization structure}
  
At a distance $r$ from the ionizing source, the number 
densities $n({\rm X}_{i}^{j})$ and $n({\rm X}_{i}^{j+1})$ of the ions 
X$_{i}^{j}$ and X$_{i}^{j+1}$   are schematically related by the 
following expression:
\begin{equation}
n({\rm X}_{i}^{j}) Q({\rm{H^{0}}}) / r^{2} K = n({\rm X}_{i}^{j+1}) n_{e} 
\alpha({\rm X}^{j}),
\end{equation}
 where $K$ is a factor taking into account the frequency distribution of 
 the ionizing radiation field and the absorption cross section (note 
 that, for simplicity, the charge exchange process is not included in 
 this equation).
Of course, ions X$_{i}^{j+1}$ can exist only if the radiation field contains 
photons able to produce these ions, and the ratio 
$n({\rm X}_{i}^{j+1})/n({\rm X}_{i}^{j})$ will be higher for higher effective 
temperatures of the ionizing source.

Integrating Eq. (1.11) over the nebular volume and using Eq. (1.2), 
it can be shown that, for a spherical nebula of constant density and 
filling factor and with an ionizing radiation of given effective temperature, the average ionic ratios are proportional to
  $(Q({\rm{H^{0}}}) n \epsilon^{2})^{1/3}$.  In other words, a nebula of density 
$n$ = $10^{4}$~\cmcub\ ionized by one star with \Tstar\ = 50\,000~K will have 
the same ionization structure as a nebula of density $n$ = 
$10^{2}$~\cmcub\ ionized by one hundred such stars.  

The ionization parameter is usually 
defined by
\begin{equation}
U =Q({\rm H^{0}}) / (4 \pi R^{2} n c ),
\end{equation}
where  $R$ is either the Str\"{o}mgren radius, or a typical 
distance from the gas cloud to the ionizing star, and 
$c$ is the speed of light. $U$ is thus directly proportional to
$(Q({\rm{H^{0}}}) n \epsilon^{2})^{1/3}$  in the case of a constant density sphere and this 
parameter describes the ionization structure.  

It is important to be aware  that equation (1.12) shows that at a given 
distance from the source, ionization drops when the density is 
increased locally (like in the case of a density clump).  On the other hand, 
of two nebulae with uniform density and ionized by the same star, 
the highest average ionization will occur for the densest one.

The presence of intense lines of low ionized species such as \Nii, 
\Sii, \Oi, is often considered in the literature as a signature of 
the presence of shocks.  Shock models indeed predict that these lines 
are strong, but it must be kept in mind that pure photoionization 
models can also produce strong low ionization lines. 
This is for example the case for nebulae 
containing regions of low ionization due to gas compression (\eg Dopita 1997
Stasi\'{n}ska \& Schaerer 1999). Another example is that of nebulae excited by very 
 high energy photons, for which the absorption 
cross-section is small and which induce a warm, only partially 
ionized zone. 
  
 \subsection{Heating and cooling}
 
 During the photoionization process, the absorption of a photon 
 creates a free electron which 
rapidly shares its energy with the other electrons present in the gas 
by elastic collisions, and thus heats the gas.  The energy gains are 
usually dominated by photoionization of hydrogen atoms, although 
photoionization of helium contributes significantly. Intuition might 
suggest that \Te\ 
will decrease away from the ionizing source, since 
the ionizing radiation field decreases because of geometrical dilution 
and absorption in the intervening layers.  This is actually not the 
case.
The total energy gains per unit volume and unit time at a distance $r$ 
from the ionizing source are schematically given by:
\begin{equation}
G =n({\rm H^{0}}) \int_{h\nu_{0}}^{\infty} (4 \pi J_{\nu}(r) / h\nu), 
a_{\nu}({\rm H^{0}}) (h\nu - h\nu_{0}) {\rm d} h\nu
\end{equation}
where 
\begin{equation}
4 \pi J_{\nu}(r) = \pi F_{\nu} (r) = 
 \pi F_{\nu}(0) (R_{\star})^{2}/r^{2} \rm{e}^{-\tau_{\nu}(r)}.
\end{equation}
If ionization equilibrium is achieved in each point of the nebula, one 
has (in the ``on-the-spot case'')
\begin{equation}
n({\rm{H^{0}}}) \int_{h\nu_{0}}^{\infty}  (4 \pi J_{\nu}(r) / h\nu)
a_{\nu}({\rm H^{0}}) d h\nu = n({\rm H^{+}}) n_{e} 
\alpha_{B}({\rm{H}}).
\end{equation}
Therefore,  $G$ can be written
\begin{equation}
G = 
n({\rm{H^{+}}}) n_{e} \alpha_{B}({\rm{H}}) <E>,
\end{equation}
where 
\begin{equation}
<E> =    \int_{h\nu_{0}}^{\infty} ( 4 \pi J_{\nu}(r) / h\nu) 
a_{\nu}({\rm H^{0}}) (h\nu - h\nu_{0}) {\rm d} h\nu
/ \int_{h\nu_{0}}^{\infty}  (4 \pi J_{\nu}(r) / h\nu) a_{\nu} ({\rm H^{0}}) {\rm d} h\nu.
\end{equation}
Thus $<E>$ can be seen as the average energy gained per photoionization, and is 
roughly independent of $r$.
It can be shown (see \eg Osterbrock 1989), that when the ionization source is a 
blackbody of temperature  \Tstar, one has $<E>  \approx (3/2) k T_{\star}$. 
Therefore:
\begin{equation}
G \propto n^{2} T_{\star} T_{e}^{-1},
\end{equation}
meaning that the energy gains are roughly proportional to the temperature of the 
ionizing stars.

Thermal losses in nebulae occur through recombination, free-free 
radiation 
and emission of collisionally excited lines. The dominant process 
is usually due to collisional excitation of ions from  heavy 
elements (with O giving the largest contribution, followed by C, N, Ne and 
S). Indeed, these ions have low-lying energy levels which can easily be 
reached at nebular temperatures. The excitation potentials of hydrogen 
lines are much higher,  so that collisional excitation of \Ho\ 
can become important only at high electron temperatures.

For the transition $l$ of ion $j$ of an element $X^{i}$, in a simple two-level 
approach and when each excitation is followed by a radiative 
deexcitation, the cooling rate can be schematically written as
\begin{equation}
L_{coll}^{ijl} = n_{e} n(X_{i}^{j}) q_{ijl} h\nu _{ijl} = 
8.63\,10^{-6}n_{e} n(X_{i}^{j}) \Omega _{ijl}/\omega  _{ijl} T_{e}^{-0.5} 
{\rm e}^{_ (\chi _{ijl}/kT_{e})} h\nu _{ijl},
\end{equation}
where $\Omega _{ijl}$ is the collision strength, $\omega _{ijl}$ is the 
statistical weight of the upper level, and $\chi _{ijl}$ is the 
excitation energy.

If the density is sufficiently high, some collisional deexcitation 
may occur and  cooling is reduced. In the two-level approach one has:
\begin{equation}
L_{coll}^{ijl} =  n_{e} n(X_{i}^{j}) n_{e} q_{ijl} h\nu _{ijl}(1/(1+n_{e} (q_{12}+q_{21})/A_{21}). 
\end{equation}

So, in a first approximation, one can write that the electron 
temperature is determined by
\begin{equation}
G = L= \sum_{ijl}^{} L_{coll}^{ijl}, 
\end{equation}
where $G$ is given by Eq. (1.18) and $L_{coll}^{ijl}$ by Eq. (1.20). 

The following properties of the electron temperature are a 
consequence of the above equations: 

-- $T_{e}$ is expected to be usually rather uniform in nebulae,
its variations are mostly determined by the mean energy of the 
absorbed stellar photons, and by the populations of the main cooling 
ions. It is only at high metallicities (over solar) that large $T_{e}$ gradients 
are expected: then cooling in the \Opp\ zone is dominated by 
collisional excitation of fine structure lines in the ground level of 
\Opp, while the absence of fine structure lines in the ground level 
of \Op\ forces the temperature to rise in the outer zones (Stasi\'{n}ska 1980a, Garnett 1992). 

-- For a given \Tstar, $T_{e}$ is generally lower at higher metallicity.

-- For a given metallicity, $T_{e}$ is generally lower for lower \Tstar.

-- For a given \Tstar\ and given metallicity, $T_{e}$ increases with density in regions 
where $n$ is larger than a critical density for 
collisional deexcitation of the most important cooling lines (around 
$5~10^{2}$ -- $10^{3}$~\cmcub).

 \subsection{Line intensities}
 
In conditions prevailing in PNe and \hii\ regions the observed 
emission lines are 
 optically thin, except for resonance lines  such as H Ly$\alpha$, 
C~{\sc iv}$\lambda$1550, N~{\sc 
v}$\lambda$1240, Mg~{\sc 
ii}$\lambda$2800, Si~{\sc iv}$\lambda$1400,  and some helium 
lines.  Also the fine structure IR lines could be optically thick in 
compact  \hii\ regions or giant  \hii\ regions (however, 
the velocity fields are generally such that this is not the case).  
The fact that most of the lines used for abundance determinations are 
optically thin makes their use robust and powerful.

The intensity ratios of recombination lines are almost independent of 
temperature. On the other hand, intensity ratios of optical and 
ultraviolet collisional lines
are strongly dependent on electron temperature if the excitation levels differ.

Abundances of metals with respect to hydrogen are mostly derived 
using the intensity ratio of collisionally excited lines with \Hb.
It is instructive to understand the dependence of such emission line ratios 
with metallicity.  Let us consider the  \Oiii/\Hb\ line ratio and 
follow its behaviour as $n$(O)/$n$(H) decreases (from now on the 
notation $n$(O)/$n$(H) will be replaced by O/H).  The temperature dependence of 
the  \Oiii\ and \Hb\ lines implies that: 
\begin{equation}
{\rm[O \,{\sc{iii}}]}\,\lambda 5007/{\rm H}\beta  
\propto n ({\rm O)} / n({\rm H}) ~ T_{e}^{0.5} {\rm e}^{-28764/T_{e}}.
\end{equation}
-- At high metallicity (O/H around $10^{-3}$ and above), cooling is 
efficient and \Te\ is low.   
Energy is mainly evacuated by the  \Oiiitoneb\ line, whose 
excitation potential is 164~K.  The cooling rate is then approximately 
given by
 \begin{equation}
L = n({\rm O^{++}}) n_{e} T_{e}^{-0.5} {\rm e}^{164/T_{e}}. 
\end{equation}

 Eq. (1.21) implies that 
\begin{equation}
{\rm[O \,{\sc{iii}}]}\,\lambda 5007/{\rm H}\beta 
\propto T_{\star} {\rm e}^{-(28764+164)/T_{e}}.  
\end{equation}
Since \Te\ increases with decreasing O/H, Eq. (1.24) shows that 
\Oiii/\Hb\ increases.  Note the value of \Oiii/\Hb\ depends on 
 \Tstar, being larger for higher effective temperatures.  
 
-- At intermediate metallicities,  (O/H of the order of $10^{-3}$ 
-- 2\,$10^{-4}$),
 cooling is still mainly due to the oxygen lines, 
 but the abundance of O/H being only moderate, \Te\ is higher, allowing collisional 
excitation of the \Oiii\ line, which now becomes the dominant coolant.  
The cooling can then be roughly expressed by:
\begin{equation}
 L = n({\rm O ^{++}}) n_{e} T_{e}^{-0.5} {\rm e}^{-28764/T_{e}}.
\end{equation}
 Eqs. (1.21) and (1.22) imply:
\begin{equation}
{\rm[O \,{\sc{iii}}]}\,\lambda 5007/{\rm H}\beta  \propto T_{\star},
\end{equation} 
i.e.  \Oiii/\Hb\ is  
proportional to \Tstar\ and independent of O/H.

-- Finally, at low metallicity, when cooling is 
dominated by recombination and collisional excitation of hydrogen, 
\Te\ becomes independent of O/H. 
From Eq. (1.22), it follows that \Oiii/\Hb\ is 
proportional to O/H. It also depends on \Tstar\ and on the average 
population of neutral hydrogen inside the nebula.

\section{Basics of abundance determinations in ionized nebulae}

 \subsection{Empirical methods}
 
 These are methods in which no check is made for the consistency of the 
 derived abundances with the observed properties of the nebulae.  They 
 can be schematically subdivided into direct methods and statistical 
 methods. 
 
  \subsubsection{Direct methods}
  
The abundance ratio of two ions is obtained from the observed intensity 
ratio of lines emitted by these ions.  For example,  
\Opp/\Hp\ can be derived from 
\begin{equation}
{\rm O ^{++}/ H^{+}} = \frac{{\rm[ O \,{\sc{iii}}]}\,\lambda 5007/{\rm 
H}\beta} 
{j_{{\rm [ O \,{\sc iii}]}(T_{e},n)}/j_{{\rm H}\beta (T_{e})}},
\end{equation}
where $j_{\rm{[O}\,{\sc iii}]}(T_{e},n)$ is the emission coefficient of the 
\Oiii\ line, which is dependent on \Te\ and $n$ (assumed uniform in the nebula). 

\Te\ can be derived using the ratio of the two 
lines \Oiiit\ and \Oiii, which have very different excitation 
potentials. Other line ratios can 
also be used as temperature indicators in nebulae, such as \rNii\ and \rSiii.  
The Balmer and Paschen jumps, 
the radio continuum and radio recombination lines also allow to estimate the 
electron temperature, but the measurements are more difficult.

The density is usually derived from intensity ratios of two lines of 
the same ion which have the same excitation energy but different 
collisional deexcitation rates.  The most common such ratio is
 \rSii. Far infrared lines can also be used to determine 
 densities. Each line pair is sensitive in a given 
density range (about 2 to 3 decades), which can be ranked as follows 
(Rubin \etal\ 1994): [N\,{\sc ii}] $\lambda$122$\mu$/205$\mu$, [O\,{\sc 
iii}] $\lambda$52$\mu$/88$\mu$, [S\,{\sc ii}] $\lambda$6731/6717, 
[O\,{\sc ii}] $\lambda$3726/3729, [S\,{\sc iii}] 
$\lambda$18.7$\mu$/33.6$\mu$, [A\,{\sc iv}] $\lambda$4740/4711, 
[Ne\,{\sc iii}] $\lambda$15.5$\mu$/36.0$\mu$, [A\,{\sc iii}] 
$\lambda$8.99$\mu$/21.8$\mu$, C\,{\sc iii}] $\lambda$1909/1907.  The 
electron density can also be measured by the ratio of high order 
hydrogen recombination lines.

Plasma diagnostic diagrams combining all the information from 
temperature- and density-sensitive line ratios can also be constructed 
for a given nebula (e.g.  Aller \& Czyzak 1983), plotting for each 
pair of diagnostic lines the curve in the (\Te,$n$) plane that 
corresponds to the observed value.  The curves usually do not 
intersect in one point, due to measurement errors and to the fact that 
the nebula is not homogeneous (and also to possible uncertainties in 
the atomic data) and provide a visual estimate of the uncertainty in 
 the adopted values of \Te\ and $n$.

The total abundance of a given element relative to hydrogen is given by the 
sum of abundances of all its ions.  In practise, not all the ions 
present in a nebula are generally observed.  The only favourable case 
is that of oxygen which in \hii\ regions is readily determined from:
\begin{equation}
\rm{O}/\rm{H} = \rm{O}^{+}/\rm{H}^{+} + \rm{O}^{++}/\rm{H}^{+}. 
\end{equation}
Note that even if \Oi\ is observed, it should not be included in the 
determination of the oxygen abundance, since the reference hydrogen line 
is emitted by H$^{+}$, while O$^{0}$ is tied to H$^{0}$.

In almost all other cases (except in some cases when multiwavelength data are 
available), one must correct for unseen ions using ionization 
correction factors.  A common way to do this in the 70' and 80' and 
even later was to rely on ionization potential considerations, which 
led to such simple expressions as: 
\begin{equation}
\rm{N}/\rm{O} = \rm{N}^{+}/\rm{O}^{+},  
\end{equation}
\begin{equation}
\rm{Ne}/\rm{O} = \rm{Ne}^{++}/\rm{O}^{++},   
\end{equation}
\begin{equation}
\rm{C}/\rm{O} = \rm{C}^{++}/\rm{O}^{++}.
\end{equation}

In high excitation PNe where \heii\ lines are seen, oxygen can be 
present in ionization stages higher than O$^{++}$.  A popular 
ionization correction scheme for oxygen 
(e.g.  Torres-Peimbert \& Peimbert 1977) was:
\begin{equation}
\frac{\rm{O}}{\rm{H}} = \frac {(\rm{He}^{+}+\rm{He}^{++})}{\rm{He}^{+}} 
\frac{(\rm{O}^{+}+\rm{O}^{++})}{\rm{H}^{+}}. 
\end{equation}
Expressions (2.29 -- 2.31) are based on the similarity the 
ionization potentials of \Cp, \Np, \Op, \Nep. Expression (2.32) is 
based on the fact that the ionization potentials of \Hep\ and 
\Opp\ are identical.

However, photoionization models show that such simple relations do not 
necessarily hold. For example, the charge transfer reaction \Opp\ + 
\Ho\ $\rightarrow$ \Op\ + \Hp\ being much more efficient than the \Nepp\ + 
\Ho\ $\rightarrow$ \Nep\ + \Hp\ one, \Nepp\ is more recombined than \Opp\ in the outer 
parts of nebulae and in zones of low ionization parameter.

Also, while it is true that no \Oppp\ ions can be found outside the 
\Hepp\ 
Str\"{o}mgren sphere, since the photons able to ionize \Opp\ are 
absorbed by \Hep, \Opp\ ions can well be present inside the \Hepp\ zone. 

Ionization correction factors based on grids of photoionization models 
of nebulae are therefore more reliable. 
Complete sets of ionization correction factors have been published 
by Mathis \& Rosa (1991) for \hii\ regions and Kingsburgh \& Barlow 
(1994) for 
planetary nebulae, or can be computed from grid of photoionization 
models such as those of Stasi\'{n}ska (1990), Gruenwald \& Viegas 
(1992) for single star \hii\ regions, 
Stasi\'{n}ska et al.  (2001) for giant  \hii\ regions, Stasi\'{n}ska et al.  (1998) for 
PNe.

However, it must be kept in mind that ionization correction factors 
from model grids may be risky too, both because the atomic physics is 
not well known yet (see Sect. 3.1) and because the density 
structure of real nebulae is more complicated than that of idealized 
models. The most robust relation seems to be 
$\rm{N}/\rm{O} = \rm{N}^{+}/\rm{O}^{+}$ (but see Stasi\'{n}ska \& 
Schaerer 1997). Such a circumstance is fortunate, given 
the importance of the N/O ratio both in \hii\ regions 
(as a constraint for chemical 
evolution studies) and in PNe (as a clue on PNe progenitors).

In spite of uncertainties, ionization correction factors
 often provide more accurate abundances than summing up ionic 
abundances obtained combining different techniques in the optical, 
ultraviolet and infrared domains. 

Note that there is no robust empirical way to correct for neutral helium to 
derive the total helium abundance. The reason is that the relative populations
of helium and hydrogen ions mostly depend on the energy distribution of the 
ionizing radiation field, while those of 
ions from heavy elements are also a function of the gas density 
distribution.

In summary, direct methods for abundance determinations are simple, powerful, and 
provide reasonable results (provided one keeps in mind the
 uncertainties involved, which will be developed in the next 
 sections). Until recently, abundances were mostly derived from 
 collisionally excited 
 optical lines. This is still the case, but the importance of  
 infrared data is growing, especially since the ISO mission. 
 IR line intensities have the advantage of being almost independent of 
 temperature. They arise from a larger variety of ions than 
 optical lines. They allow to probe regions highly obscured by dust. 
 However, they suffer from beamsize and calibration problems which 
 are far more difficult to overcome than in the case of optical 
 spectra. 
 Abundance determinations using recombination lines of heavy 
 elements have regained interest these last years. They require high 
 signal-to-noise spectroscopy since the strengths 
 of recombination lines from heavy elements are typically 0.1\% of those 
 of hydrogen Balmer lines. They will be discussed more thoroughly in 
 the next sections, since they unexpectedly pose one of the major
 problems in nebular astrophysics.

 \subsubsection{Strong line or statistical methods}
  
  When the electron temperature cannot be determined, for example 
because the observations do not cover the appropriate spectral range 
or because temperature sensitive lines such as \Oiiit\ cannot be 
observed, one has to go for statistical methods or ``strong line 
methods''.  These methods have first been introduced by Pagel et al. 
(1979) to derive metallicities in giant extragalactic \hii\ regions.  
They have since then being reconsidered and recalibrated by many authors,
among which Skillman (1989), McGaugh (1991, 1994), Pilyugin 
(2000, 2001).  

Pagel et al. (1979) proposed to use 
the 4 strongest lines of O and H : \Ha, \Hb, \Oii\ and 
\Oiii.  
From Sect. 1, the main parameters governing the relative intensities of 
the emission lines in a nebula are : $<$\Tstar $>$, the mean effective 
temperature of the ionization source, the gas density distribution 
(parametrized by $U$ in the case of homogeneous spheres), and the 
metallicity, represented by O/H.
 Luckily oxygen is at the same time the main coolant in nebulae, and 
 the element whose abundance is most straightforwardly related to the 
 chemical evolution of galaxies.   
The 
spectra must be corrected for reddening, which is done by comparing 
the observed \Ha/\Hb\ ratio with the case B recombination value at a 
typical $T_{e}$  and assuming a 
reddening law (see Sect. 3.3).
Therefore two independent line 
ratios, \Oii/\Hb\ and \Oiii/\Hb, remain to determine three quantities.  
Statistical methods rely on the assumption that $<$\Tstar $>$ (and 
possibly $U$) are closely linked to the metallicity, and that 
it is the metallicity which drives the observed line ratios.  Basing 
on available photoionization model grids, Pagel et al. 
showed that (\Oii\ + \Oiii)/\Hb, later called 
O$_{23}$,  could be used as an indicator of  O/H at metallicities above 
half-solar. Skillman (1989) later argued that this ratio could also 
be used in the low metallicity regime, in cases when the observations did not 
have sufficient signal-to-noise to measure the \Oiiit\ line intensity. 
McGaugh (1994) improved the method and proposed to use both \Oiii/\Oii\ and O$_{23}$ to 
determine simultaneously O/H and $U$ (his method should perhaps be 
called the O$_{23+}$ method).  For the reasons explained above, 
the same value of (\Oii\ + \Oiii)/\Hb\  can correspond to either a high or 
a low value of the metallicity. A useful discriminator is \Nii/\Oii, 
since it is an empirical fact that  \Nii/\Oii\ 
increases with O/H (McGaugh 1994).

The expected accuracy of statistical methods is typically 0.2 -- 0.3 dex, 
the method being particularly insensitive in the turnover region at 
O/H around  $3~10^{-4}$. 

On the low metallicity side, the method can easily be calibrated with 
 data on metal-poor extragalactic \hii\ regions where the 
\Oiiit\ line can be measured.  Recently, 
Pilyugin (2000) has done this using the large set of excellent quality observations of
 blue compact galaxies by Izotov and coworkers 
(actually, the strong line method proposed by Pilyugin differs 
somewhat from the O$_{23}$ method, but it relies on similar principles). 
 He showed that the method works extremely well at low metallicities 
 (with an accuracy of about 0.04~dex). 
 This is a priori surprising, since giant \hii\ 
regions are powered by clusters of stars that were formed almost 
coevally.  The most massive stars die gradually, 
inducing a softening of the ionizing radiation 
field on timescales of several Myr, which should affect the O$_{23}$ 
ratio, as shown by McGaugh (1991) or Stasi\'{n}ska (1998).  As discussed by Stasi\'{n}ska  
et al. (2001), data on \hii\ regions in blue compact dwarf galaxies are 
probably biased towards the most recent starbursts, and the dispersion 
in $<$\Tstar$>$  is not as large as could be expected a priori. 
Another possibility, advocated by Bresolin et al. (1999) in their study 
of giant \hii\ regions in 
spiral galaxies is that some mechanism must disrupt the 
 \hii\ regions after a few Myr. Of course, the  O$_{23}$ method is 
expected of much lower accuracy when applied to \hii\ regions ionized 
by only a few stars, since in that case the ionizing radiation field 
varies strongly from object to object.

On the high metallicity side (O/H larger than about $5~10^{-4}$), 
the situation is much more complex.  In 
this regime, there is so far no direct determination of O/H to allow a 
calibration of the O$_{23}$ method since the \Oiiit\ line is too weak to be 
measured (at least with 4~m class telescopes).  The calibrations 
rely purely on models but it is not 
known how well these models represent real \hii\ regions.  Besides, at these 
abundances, the  \Oii\ and 
\Oiii\ line intensities are extremely sensitive to 
any change in the nebular properties (Oey \& Kennicutt 1993, Henry 1993, Shields 
\& Kennicutt 1995).  
Note that the calibration proposed by Pilyugin (2001) of his related  
 X$_{23}$  method in 
the high metallicity regime actually refers to O/H ratios that are 
lower than  5 $10^{-4}$.

Other methods have been proposed as substitutes to the O$_{23}$ 
method. The S$_{23}$ method, proposed by V\'{\i}lchez \& Esteban (1996) and 
D\'{\i}az \& P\'{e}rez-Montero (2000) relies on the same principles as the O$_{23}$ 
method, but uses  ([S~{\sc ii}] $\lambda$6716, $\lambda$6731 + [S~{\sc iii}] 
$\lambda$9069, $\lambda$9532)/\Hb\ (S$_{23}$) instead of  (\Oii\ + \Oiii)/\Hb. 
One advantage over the O$_{23}$ method is that the relevant line 
ratios are less affected by reddening. Besides, the excitation levels 
of the \Sii\ and \Siii\ lines are lower than those of the \Oii\  and 
\Oiii\ lines, so that S$_{23}$ increases with metallicity in a wider 
range of metallicities than O$_{23}$ (the turnover region for  S$_{23}$ is expected at
 O/H around $10^{-3}$). Unfortunately, \Siii\ is more difficult to 
 observe than \Oiii. Oey \& Shields (2000) argue that the  S$_{23}$ 
 method is more sensitive to $U$ than claimed by Diaz \& Perez-Montero 
 (2000). This would require futher checks, but in any case, the S$_{23}$ 
 method could be refined into an S$_{23+}$ method in the same way as 
 the O$_{23}$ was refined into the O$_{23+}$ method. 
 
Stevenson et al. (1993) proposed to use [Ar~{\sc iii}] 
$\lambda$7136] / [S~{\sc iii}] $\lambda$9532 as 
an indicator of the electron temperature in metal-rich \hii\ regions, 
and therefore of their metallicity. This method relies on the idea
 that the Ar/S ratio is not expected to vary significantly from object 
 to object, and that the \Arpp\ and \Spp\ zones should be coextensive. 
 However, photoionization models show that, because of the strong 
 temperature gradients expected at high metallicity, this method could lead to 
 spurious results. 
 
 Alloin et al. (1979)  proposed to use 
\Oiii/\Nii\  as a statistical metallicity indicator. While this line 
 ratio depends on an additional parameter, namely N/O, the accuracy of 
 this method turns out to be similar to that of statistical 
 methods mentioned above. More recently, Storchi-Bergman et al. 
 (1994), van Zee et al. (1998) and Denicolo et al. (2001) advocated for 
 the use of
 the \Nii/\Hb\ ratio (N$_{2}$) as metallicity indicator. Similarly 
 to \Nii/\Oiii, this 
 ratio  shows to be correlated with O/H over the entire range of 
 observed metallicities in
 giant \hii\ regions. The reason why, contrary to the  O$_{23}$ 
 ratio, it increases with O/H even at high metallicity is due to a 
 conjunction of 
 \Nii/\Hb\ being less dependent on 
 \Te\ than O$_{23}$, N/O being observed to increase with O/H in giant 
 \hii\ regions (at high 
 metallicity at least) and $U$ tending to decrease with metallicity. 
 The advantage of this ratio is that it is 
 independent of reddening and of flux calibration, and is only weakly
 affected by underlying stellar absorption in the case of observations 
 encompassing old stellar populations. This makes it extremely 
 valuable for ranking metallicities of galaxies up to redshifts about 
 2.5. 
 
 As mentioned above, statistical methods for abundance determinations 
 assume that the nebulae under study form a one parameter family. This is why they work 
 reasonably well in giant \hii\ regions. They are not expected to 
 work in planetary nebulae, where the effective temperatures range 
 between 20\,000~K and 200\,000~K. Still, it has been shown 
 empirically that there is an upper envelope in the \Oiii/\Hb\ vs. O/H 
 relation (Richer 1993), probably corresponding to PNe with the 
 hottest central stars. The existence of such an envelope can be used 
 to obtain lower limits of O/H in PNe located in distant galaxies. 
 
 \subsection{Model fitting}
 
  \subsubsection{Philosophy of model fitting}

A widely spread opinion is that 
photoionization model fitting provides the most accurate abundances.  This 
would be true if the constraints were sufficiently numerous (not only on emission line ratios,
but also on the stellar 
content and on the nebular gas distribution) and if the model fit 
were perfect (with a photoionization code treating correctly all the 
relevant physical processes and using accurate atomic data).
These conditions are  never met in practise, and it is therefore worth 
thinking, before embarking on a detailed photoionization modelling, 
what is the aim one is pursueing.

Two opposite situations may arise when trying to fit  observations 
with a model. 

The first one occurs  when the number of strong 
constraints is not sufficient, especially when no direct \Te\ indicator 
is available. Then various models may be equally well compatible
with the observations. For example, from a photoionization model 
analysis Ratag et al. (1997) 
derive an O/H ratio of 2.2~$10^{-4}$ for the PN M 2-5. 
However, if one explores the range 
 of acceptable photoionization models one finds 
 two families of solutions (see Stasi\'{n}ska 2002). The first has 
O/H $\simeq$ 2.4~$10^{-4}$, the second has O/H $\simeq$ 1.2~$10^{-3}$! 
The reason for such a double solution is simply the behaviour of 
\Oiii/\Hb\ or \Oii/\Hb\ with metallicity, 
as explained in Sect. 1.3. Note that both families of models
reproduce not only the observed 
line ratios  (including upper 
limits on unobserved lines) but also the 
nebular size and total \Hb\ flux.

The other situation is when, on the contrary, one cannot find any solution that reproduces at 
the same time the \rOiii\ line ratio and the constraints of the 
distribution of the gas and ionizing star(s) (e.g. Pe\~{n}a et al. 
1998, Luridiana et al. 1999, Stasi\'{n}ska \& Schaerer 1999).  
 The model that best reproduces the strong oxygen lines has a 
 different value of O/H than would be derived using an empirical  
electron-temperature based  method.   The difference 
between the two can amount to factors as large as 2 (Luridiana et al.
1999). It is difficult to say a 
priori which of the two values of O/H -- if any -- is the correct one. 

The situation where the number of strong constraints is large and 
everything is satisfactorily fitted with a 
photoionization model is extremely rare. One such example is the case 
of the two PNe in the Sgr B2 galaxy, for which high signal-to-noise 
integrated spectra are available providing several electron temperature and 
density indicators with accuracy of a few \%. Dudziak et al. (2000) reproduced 
the 33  (resp. 27) 
independent observables (including imagery and photometry) with 
two-density component models having 18 (resp. 14) free parameters for Wray 
16-423 (resp. He 2-436). Still, the models are not really unique. The 
authors make the point that they can reproduce the 
 present observations with a range of values for 
 C/H and  \Tstar. Yet, the derived abundances 
 are not significantly different from those obtained from the same 
 observational data by Walsh et al. 
 (1997) using the  empirical  method. The only notable difference is 
 for sulfur whose abundance from the models is larger by 50\%, and for
 nitrogen whose abundance from the models is larger by a factor of 2.8 
 in the case of He 2-436. This apparent 
 discrepancy for the nitrogen 
 abundance actually disappears if realistic error bars are 
 considered for the direct abundance determinations (rather than the error bars quoted in the papers). Indeed, the fact 
 that the nebular gas is rather dense, with different density indicators 
 pointing at 
 densities from 3~$10^{3}$\cmcub\ up to over $10^{5}$\cmcub\ 
 introduces important uncertainties in the temperature derived from 
 \rNii\ due to collisional deexcitation. It must be noted that 
 realistic error 
 bars on abundances derived from model fitting are extremely difficult 
 to obtain, since this would imply the construction of a tremendous number 
 of models, all fitting the data within the observational errors.
 
 To summarize, abundances are not necessarily better determined from 
 model fitting. However, model fitting, if done with a sufficient 
 number of constraints, provides 
 ionization correction factors relevant for the object under study 
 that should be more accurate than simple formulae derived from grids of 
 photoionization 
 models.  This could be called a ``hybrid method'' to derive abundances. 
 Such a method was for example used by  Aller \& Czyzak (1983) and 
 Aller \& Keyes (1987) to derive the abundances in a large sample of 
 Galactic planetary nebulae, and is still being used by Aller and his 
 coworkers. It must however be kept in mind that if photoionization 
 models do not reproduce the temperature sensitive line ratios, this 
 actually points to a problem that has to be solved before one can 
 claim to have obtained reliable abundances. 
 
 Ab initio photoionization models are sometimes used to estimate 
 uncertainties that can be expected in abundance determinations 
 from empirical methods. For example  
 Alexander \& Balick (1997) and Gruenwald \& Viegas (1998)
 explored the validity of traditional ionization 
 correction factors in the case of spatially resolved observations. A 
 complete discussion of uncertainties should also take into account 
 uncertainties in the atomic data and the effect of a simplified 
 representation of reality by photoionization models.

\subsubsection{Photoionization codes}

Photoionization codes are built to take into account all the major physical processes that govern 
the ionization and temperature structure of nebulae.  In addition to 
photoionization, recombination, free-free radiation, collisional 
excitation they consider collisional ionization (this is 
important only in regions of coronal temperatures), charge exchange 
reactions, which are actually a non negligible cause of recombination 
for heavy elements, especially if the physical conditions are such 
that the population of residual hydrogen atoms in the ionized gas 
exceeds $10^{-3}$.  Some codes are designed to study nebulae that are not 
in equilibrium and they may include such processes as mechanical 
heating and expansion cooling.

Most nebular studies use static photoionization codes, which assume 
that the gas is in ionization and thermal equilibrium.
The most popular one is CLOUDY developed by Ferland and co-workers, for which an 
extensive documentation is available and which is widely in use (see Ferland 1998, and 
http://www.pa.uky.edu/~gary/cloudy/ for the latest release).
Several dozens of independent photoionization codes suited for the 
study of PNe and \hii\ regions have been constructed over the years. 
Some of them have been intercompared at several workshops 
(P\'{e}quignot 1986, Ferland et al. 1996 and Ferland \& Savin 2001). 
The codes mainly differ in the 
numerical treatment of the transfer of the ionizing photons produced 
in the nebula: on the spot reabsorption, outward-only approximation (most 
codes presently), full treatment (either with classical techniques as 
in Rubin 1968 or Harrington 1968  or with Monte-Carlo techniques as in Och 
et al. 1998). They also differ in their 
capacity of handling different geometries. Most codes are built in 
plane parallel or spherical approximations, but a few are built in 3D 
(Gruenwald et al. 1997, Och et al. 1998). 
While 3D codes are better suited to 
represent the density distribution in real nebulae, their use
is hampered by the fact that the number of free parameters is 
extremely large. Presently, simpler codes are usually sufficient to pinpoint 
difficulties in fitting observed nebulae  within our present 
knowledge of the physical processes occuring in them
and to settle error bars on abundance determinations.

When the timescale of stellar evolution becomes comparable to the 
timescale of recombination processes, the assumption of ionization 
equilibrium is no more valid.  This for example occurs in PNe with 
massive ( $>$0.64\Ms) nuclei, whose temperature and luminosity drop in
a few hundred years while they evolve towards the white 
dwarf stage.  In that case, the real ionization state of the gas is 
higher than would be predicted by a static photoionization model, and 
a recombining halo can appear.  To deal with such situations, one needs 
time dependent photoionization codes, such as those of Tylenda (1979), or Marten 
\& Szczerba (1997).

The nebular gas is actually shaped by the dynamical effect of the 
stellar winds from the ionizing stars. This induces shocks that 
produce strong collisional heating at the ionization front 
or at the interface between the 
main nebular shell of swept-up gas and the hot stellar wind bubble.  
On the other hand, expansion contributes to the cooling of the nebular gas. 
Several codes have been designed to treat simultaneously the 
hydrodynamical equations and the microphysical processes either in 
1D (e.g. Schmidt-Voigt \& K\"{o}ppen 1987a and b , Marten \& 
Sch\"{o}nberner 1991,  Frank \& Mellema 
1994a, Rodriguez-Gaspar \& Tenorio-Tagle 1998) or in 
2D  (Frank \& Mellema 1994b, Mellema \& Frank 1995, Mellema 1995).
It may be that some of the problems found with static codes will find 
their solution with a proper dynamical description.  However, so far, 
for computational reasons, the microphysics and transfer of radiation 
is introduced in a more simplified way in these codes.  Also, it is 
much more difficult to investigate a given problem with 
such codes, since the present state of an object is the result of its 
entire history, which has to be modelled ab initio.

\section{Main problems and uncertainties in abundance determinations}

The validity of derived abundances  depends on the 
quality of the data and on the method of analysis.  
Typical quoted values for the uncertainties are 0.1 -- 0.25 dex for 
ratios such as O/H, N/H, Ne/H, a little more for S/H, A/H, C/H, a 
little less for N/O, Ne/O and a few \% for He. 
The optimism of the investigator is an important factor in the 
evaluation of the accuracy.
This section comments on the various sources of uncertainties in 
abundance determinations.

\subsection{Atomic data}

Reviews on atomic data for abundance analysis have been given by 
Mendoza (1983), Butler (1993), Storey (1997), Nahar (2002). On-line 
atomic data bases are available from different sites. For example 
 http://plasma-gate.weizmann.ac.il/DBfAPP.html  provides links to many sites 
of interest, including the site of CLOUDY. The XSTAR 
atomic data base, constructed by Bautista \& 
Kallman (2001) and used in the photoionization code XSTAR can be found 
at http://heasarc.gsfc.nasa.gov/docs/software/xstar/xstar.html.

The OPACITY and IRON projects (Seaton 1987, Hummer et al. 1993) have 
considerably increased the reliability of atomic data used for 
nebular analysis in the recent years. 
In the following, we simply raise a few important points.

	\subsubsection{Ionization, recombination and charge exchange}
	
Until recently, photoionization cross sections and recombination 
(radiative and dielectronic)
coefficient sets used in photoionization computations were not 
obtained self-consistently. 
Photoionization and recombination calculations are presently being 
carried out using the same set of eigenfunctions as in the IRON 
project (Nahar \& Pradhan 1997, Nahar et al. 2000).   The 
expected overall uncertainty is 10 -- 20\%. Experimental checks on a 
few species (see e.g. Savin 1999) can provide benchmarks for 
confrontation with numerical computations. 

Concerning charge exchange, only a few detailed computations are 
available (see references in the compilation by Kingdon \& Ferland 
1996). Coefficients computed with the Landau-Zeener approximation are 
available for most ions of interest. They are unfortunately rather 
uncertain. Differences with 
coefficients from quantal computations, which are available for a few species 
only, can be as large as a factor 3. 

Due to the uncertainties in atomic parameters, the ionization structure 
predicted by photoionization models is so far expected to be accurate only 
for elements from the first and second row of the Mendeleev table. 

\subsubsection{Transition probabilities, collision strengths and 
effective recombination coefficients}
	
	The atomic data to compute the emissivities of optical forbidden lines 
have been recently recomputed in the frame of the IRON 
project (Hummer et al. 1993). 
The expected accuracies are typically of 10\% for second row elements, 
however, the uncertainty is difficult to determine internally. 
Comparison with laboratory data is scarce, and actually, PNe are 
sometimes used as laboratories to test atomic physics 
calculations. For example, van Hoof et al. (2000) studied 3 PNe in 
detail and concluded that the \nev\ collision strenghts computed by
Lennon \& Burke 
(1994) should be correct within 30~\%, contrary to previous 
suggestions by Oliva et al. (1996) and Clegg et al. (1987).  
  Another example is the density derived from 
  [O\,{\sc{iii}}]\,$\lambda$52\,$\mu$m/[O\,{\sc{iii}}]\,$\lambda$88\,$\mu$m, 
  which is 
  significantly lower than derived from \rSii\ and \Ariv\ for a large 
  sample of PNe observed by ISO (Liu et al. 2001).  These authors
  argue that \oiii\ IR lines can be emitted from rather low 
density components but it could just be that the atomic data are in 
error. 

Concerning recombination lines, the effective recombination rates for 
lines from 
hydrogenic ions have been recomputed by Storey \& Hummer (1995) 
and by Smits (1996) for \Hei. For C, N, O, estimates for all 
important optical and UV transitions are given   
by P\'{e}quignot et al. (1991).
Detailed computations of effective recombination 
coefficients 
are now available for lines from several ions of C, N, O and Ne (see 
e.g. a compilation in Liu et al. 2000). Note however that these do 
not include dielectronic recombination for states with high quantum 
number, which may have important consequences for the interpretation 
of recombination line data (see Sect. 3.6)

 \subsection{Stellar atmospheres}
 
 The ionization structure of nebulae obviously depends on the spectral 
 distribution of the 
stellar radiation field.
The theory of stellar atmospheres has made enormous progress these 
last years, due to advanced computing facilities. Several sets of 
models for massive O stars and for PNe nuclei are now available. 
The most detailed stellar atmosphere computations 
now include non-LTE effects and blanketing for numerous elements 
(e.g.   Dreizler \& Werner 1993, Hubeny \& Lanz 1995, Rauch et al. 2000) 
and supersede previous works.  
The effect of winds, which is especially important for evolved stars 
such as Wolf-Rayet stars, is included in several codes, although 
with different assumptions (Schaerer \& de Koter 1997, 
Hillier \& Miller 1998, Koesterke et al. 2000, Pauldrach et al. 2001). 

The resulting  model atmospheres differ 
considerably between each other in the extreme UV. This has a 
strong impact on the predicted nebular ionization structure (see e.g. 
Stasi\'{n}ska \& Schaerer 1997 for the Ne and  the \Np/\Op problems). 
Actually, the confrontation of photoionization models with 
observations of nebulae is expected to provide tests of the ionizing 
fluxes from model 
atmospheres  (see  Oey et al. 2000, Schaerer 
2000, Giveon et al. 2002, Morisset et al. 2002). This is especially 
rewarding with the ISO data which 
provide accurate measurements for  many fine-structure lines of adjacent ions.

For exploration purposes, it is sometimes sufficient to assume that the 
ionizing stars radiate as blackbodies, e.g.
 when interested in a general description of
the temporal evolution of PNe spectra as their nuclei evolve from the AGB 
to the white 
dwarf stage (e.g. Schmidt Voigt \& K\"{o}ppen 1987a, b,  
Stasi\'{n}ska et al. 1998). On the other hand, for a detailed model 
analysis of specific objects, the black body approximation
is generally not well suited. For example, the emission of \nev\ lines 
in PNe cannot be understood when using blackbodies of reasonable 
temperatures.

 \subsection{Reddening correction}
 
 The usual dereddening procedure is to derive the logarithmic 
 extinction at \Hb, $C$, from the observed \Ha/\Hb\ ratio, assuming 
 that the intrinsic one has the value (\Ha/\Hb)$_{\rm{B}}$ predicted by 
  case B recombination:
 \begin{equation} 
C = [ {\rm log} ({\rm H}\alpha / {\rm H}\beta)_{\rm B} - {\rm log} 
({\rm H}\alpha  / 
{\rm H}\beta)_{obs} ] / (f_{\alpha} - f_{\beta}), 
\end{equation}
where $f_{\alpha}$ and $f_{\beta}$ represent the values of the reddening 
law at the wavelengths of the \Ha\ and \Hb\ lines respectively.

Then, for any observed line ratio ($F_{\lambda 1} / F_{\lambda 2})_{obs}$ 
one can obtain the reddening corrected value ($F_{\lambda 1} / F_{\lambda 
2})_{corr}$ 
from: 
 \begin{equation} 
{\rm log} (F_{\lambda 1} / F_{\lambda 2})_{corr} = 
{\rm log} (F_{\lambda 1} / F_{\lambda 2})_{obs}
+ C (f_{\lambda 1} - f_{\lambda 2}). 
\end{equation}

Ideally, one can iterate after having determined the electron 
temperature of the plasma, to use a value of
(\Ha/\Hb)$_{\rm{B}}$ at the appropriate temperature.

There are nevertheless several problems. One is that the 
extinction law is not universal. As 
shown by Cardelli et al. (1989), it depends on the parameter  $R_{V}=A_{V}/E(B-V)$, 
where  $A_{V}$ is the absolute extinction in $V$ and $E(B-V)$ is the 
color excess. While the canonical value of $R_{V}$ is 3 -- 3.2, the 
actual values range from 2.5 to 5 (Cardelli et al. 1989, Barbaro 
et al. 2001, Patriarchi et al. 2001). Objects located in the Galactic 
bulge suffer from an extinction characterized by a low value of 
$R_{V}$ (e.g. Stasi\'{n}ska et al. 1992, Liu et al. 2001). 
Cardelli et. el. (1989) attribute these differences in extinction laws 
between small and large values of $R_{V}$ to the presence of 
 systematically larger particles in dense regions. These 
variations in $R_{V}$ have a significant effect on line ratios when 
dealing with ultraviolet spectra. It is therefore convenient to 
link the optical and ultraviolet spectra by using line ratios 
with known intrinsic value, such as 
He~{\sc ii} $\lambda$1640 / He~{\sc ii} $\lambda$4686. 

Another difficulty is that dust is not necessarily entirely located between the 
object and the observer as in the case of stars. Some extinction may 
be due to dust  mixed 
with the emitting gas. In that case, the wavelength dependence of the 
extinction is different and strongly geometry dependent (Mathis 
1983). One way to proceed, which alleviates this problem, is to use 
the entire  set of observed hydrogen lines and fit their ratios to the 
theoretical value, which then gives an empirical reddening law to 
deredden the other emission lines. This, however, is still not perfect, 
since the extinction suffered by lines emitted only at the 
periphery of the nebula, or, on the contrary, only in the central 
parts, is different from the extinction suffered by hydrogen lines 
which are emitted in the entire nebular body. 
The problem is further complicated by scattering effects (see e. g. 
Henney 1998).

In the case of giant \hii\  regions, where the observing slit 
encompasses  stellar light, one must first correct for the stellar 
absorption in the hydrogen lines. This can be done in an iterative 
procedure, as outlined for example by Izotov et al. (1994).  

A further problem is that the intrinsic hydrogen line ratios may  deviate 
from case B theory. This occurs for example in nebulae with high 
electron temperature ($\sim$ 20\,000~K), where collisional 
contribution to the emissivity of the lowest order Balmer lines may 
become significant. In that case, a line ratio corrected assuming case 
B for the hydrogen lines, $(F_{\lambda 1} / F_{\lambda 2})_{\rm{B}}$ 
is related to the true line ratio $(F_{\lambda 1} / F_{\lambda 
2})_{true}$ by:  
\begin{equation}
{\rm log} (F_{\lambda 1} / F_{\lambda 2})_{\rm{B}} -
{\rm log} (F_{\lambda 1} / F_{\lambda 2})_{true} 
= [{\rm log} ({\rm H}\alpha / {\rm H}\beta)_{\rm{B}} -
 {\rm log} ({\rm H}\alpha  / 
{\rm H}\beta)_{true} ] (f_{\lambda 1} - f_{\lambda 2}) / (f_{\alpha} - f_{\beta}).
\end{equation}
The error is {\em independent of the real extinction} and can be 
large for $\lambda 1$ very different from $\lambda 2$. For example, 
it can easily reach a factor 1.5 -- 2 for \Ciii/\Oiii\ (see 
Stasi\'{n}ska 2002).

Whatever dereddening procedure is adopted, it is good practise to check whether 
the H$\gamma$ /H$\beta$ value has the expected value. If not, the
\rOiii\ ratio will be in error by a 
similar amount. 

 \subsection{Aperture correction, nebular geometry and density 
 inhomogeneities}
 
Observations are made with apertures or slits that often have a 
smaller projected size on the sky than the objects under study.  
When combining data obtained with different instruments, one needs 
to correct for aperture effects.  To merge spectra obtained by IUE 
with optical spectra, 
one can use pairs of  lines of the same ion
such as He~{\sc ii} $\lambda$1640 and He~{\sc ii} $\lambda$4686.  
However, ionization stratification and reddening
make the problem difficult to solve. One can also use
C~{\sc iii}] $\lambda$1909 and C~{\sc ii} $\lambda$4267, but this 
involves additional difficulties (see Sect. 3.6). Summarizing, 
aperture corrections can  be wrong by a factor as large as 2 (Kwitter 
\& Henry 1998, van Hoof et al. 2000).

Interpretation of emission line ratios should care whether the 
 observing slit covers the entire nebula, at 
least in the estimation of error bars on derived quantities. This is 
especially important when the observations cover only a 
small fraction of the total volume. Gruenwald \& Viegas (1992)
 have published line of sight results for grids of  \hii\ region 
 models, that can be used to estimate the ionization correction 
 factors relevant to \hii\ region spectra observed with small apertures. 
  Alexander \& Balick (1997) and Gruenwald \& Viegas (1998) 
  have considered the case of PNe, 
  and shown that traditional ionization correction factors
may strongly overestimate (or underestimate) the N/H ratio in the case 
when the slit size is much smaller than the apparent size of the nebula. 
The ratio N/O is less affected by line of sight effects. 
The problem is of course even worse in real nebulae than in those 
idealized models, due to the presence of small scale density 
variations. Integrated spectra have the merit on being less dependent 
on local conditions and of being more easily comparable to models. 
For extended nebulae, they can be obtained by scanning the slit 
across the face of the nebula (van 
Hoof et al. 2000, Liu et al. 2000), or by using specially designed 
nebular spectrophotometers (Caplan et al. 2000).

Tailored modelling taking explicitly into account 
departure from spherical symmetry is still in its infancy. One may 
mention the work of 
 Monteiro et al. (2000) who  constructed a 3D photoionization model to 
reproduce the narrow band HST images and velocity profiles of the PN NGC 
3132 and concluded that 
 this nebula has a diabolo shape despite its elliptical appearance. 
 For the abundance determination however, which is the topic of this review, 
 their finding has actually  
 no real incidence. 
  
More relevant for abundance determinations are the works of 
Sankrit \& Hester (2000) and Moore et al. (2000), who modelled 
individual filaments in large nebulae, trying to reproduce the 
emission line profiles in several lines. Such a method uses many 
more constraints than classical \Te-based methods to 
derive abundances, but would need additional line 
ratios, and especially the \Te\ indicators, to be validated.

If large density contrasts occur in ionized nebulae, 
the use of forbidden lines for abundance determinations may induce 
some bias if collisional deexcitation is important. 
These biases have been explored by Rubin (1989) and his ``maximum bias 
table'' can be used to confine errors in abundances due to these 
effects.  

 \subsection{Spatial temperature variations}
 
	\subsubsection{Temperature gradients}
	
At high metallicities, as explained above, large temperature gradients are 
expected in ionized nebulae.  Therefore, empirical methods based on 
\rOiii\  will underestimate the abundances of heavy elements, since the 
 \Oiiit\ line will be essentially emitted in the high temperature 
 zones, inducing a strong overestimate of the average \Te. 
Therefore, although with very large telescopes it will now be possible 
to measure \Oiiit\ even in high metallicity giant \hii\ regions, one should 
refrain from exploiting this line in the usual way. Doing this, one would 
necessarily find sub-solar oxygen abundances, even for giant 
\hii\ regions with metallicities well above solar (Fig. 3 of Stasi\'{n}ska 2002).
High metallicity \emph {luminous} PNe offer a much safer way to  probe 
 the metallicity in central parts of galaxies 
 (see Sect. 5.3 for the relevance of PNe 
as  metallicity indicators of their environment). Indeed the higher 
effective temperatures and the higher densities in \emph {luminous} PNe 
induce higher values of \Te\ in the \Opp\ zone and a shallower 
temperature gradient,  leading to a negligible bias in the derived 
abundances (see
 Fig. 4 of Stasi\'{n}ska 2002).  

While \Te -based empirical methods are biased for metal rich 
giant \hii\ regions, tailored photoionization modeling to reproduce the 
\emph {distribution} of the emission in the \Ha, \Hb, \Hei, \Oii and 
\Oiii\ lines are worth trying.  As suggested by 
Stasi\'{n}ska (1980a), at high metallicity, regions emitting strongly  
\Oiii\ will be decoupled from 
the regions emitting strongly in the recombination lines, 
and would be almost cospatial with regions emitting most of \Oii. 
 While the models of Stasi\'{n}ska (1980a)  were made under spherical 
symmetry, the statement is more general, because it relies of the 
principles of ionization and thermal balance outlined in Sects. 1.1 
and 1.2.  

	\subsubsection{Small scale temperature variations}

If the temperature in a nebula is not uniform, \Te -based 
empirical abundances are biased.  Peimbert (1967) 
developed a mathematical formulation to evaluate the bias. It is based 
on the Taylor expansion of the average temperature
\begin {equation} 
T_{o}(N_{i})=\frac{\int{T_{e}N_{i}n_{e}{\rm d}V}}{\int{N_{i}n_{e}{\rm d}V}} 
\end {equation} 
 defined for each ion $N_{i}$, using the
{\em r.m.s.} temperature fluctuation 
\begin {equation} 
t^{2}(N_{i})= 
\frac{\int{(T_{e}-T_{o}(N_{i}))^{2}N_{i}n_{e}{\rm d}V}} 
{{T_{o}(N_{i})}^{2}\int{N_{i}n_{e}{\rm d}V}}. 
\end {equation}

From comparison of temperatures 
measured by different methods, this temperature 
fluctuation scheme led to conclude that  temperature fluctuations 
are common in nebulae, with typical values of $t^{2}$ = 0.03 -- 0.05 
(see references in Peimbert 1996, Stasi\'{n}ska 1998, Mathis et al. 
1998, Esteban 2002). The case is not always easy to make: 
the determination of 
the continuum in the vicinity of the Balmer jump is difficult, the combination of 
data from different instruments for the comparison of far 
infrared data with optical ones involves many potential sources of 
errors, lines of \Opp\ and of H  are not emitted in 
coextensive zones etc ...  
Nevertheless, the observational results seem overwhelming. And, as noted by 
Peimbert (2002), the value of $t^{2}$ found in such a way is never negative!

Note that in the PNe NGC 6153, NGC 7009, M1-42 and M2-36, (Liu et al. 1995, 
2000, 2001,  Luo et al. 2001) much larger values of $t^{2}$, of the order of 
0.1, would be derived from the comparison of optical recombination 
lines (ORL) to collisionally excited lines (CEL). But this may 
be another problem (see Sect. 3.6).

 A value of $t^{2}\sim 0.04$, in the scheme of Peimbert (1967), 
 typically leads to an 
 underestimation of O/H by about 0.3 dex \footnote {N/O and 
 Ne/O ratios are less affected by temperature fluctuations than  N/H, 
 since N/O and Ne/O abundance determinations rely on lines with similar 
 temperature dependences and emitted in roughly the same zones.}.  
 It is thus extremely important to determine 
 whether temperature fluctuations exist or whether they are an
 artefact of the techniques employed.  And, if they 
 really exist, to understand their nature and possibly derive some 
 systematics to account for them in abundance derivations.  Note that, 
 so far, the evidences are always \emph {indirect}, based on the 
 comparison of different methods to estimate \Te. Only
  mapping the nebulae with appropriate 
 sensitivity and spatial resolution in the temperature diagnostic 
 lines could give \emph {direct} evidence of small 
 scale fluctuations.  In the planetary nebula NGC 6543, HST mapping of 
 \rOiii\  shows much smaller spatial temperature variations than expected 
 for this object from indirect measurements (Lame et al. 1998).  
 In NGC 4361, long slit spectroscopy gives a \emph {surface} temperature 
 fluctuation
 $t^{2}_{s}\sim 0.002$ (Liu 1998).  In Orion, long slit mapping of the 
 Balmer decrement gives  $t^{2}_{s}\sim 0.001$ 
 (Liu et al.  1995a).  All 
 these observed $t^{2}_{s}$ translate into \emph {volume} temperature 
 fluctuations $t^{2}\leq$ 0.01.

Actually, the value of $t^{2}$ defined by Eq. (3.37)
is not strictly speaking equal to the value of $t^{2}$ derived 
observationally, for 
example from the comparison of temperatures derived from \rOiii\  and 
from the Balmer jump. Kingdon \& Ferland (1995) introduced the 
notation $t^{2}_{str}$ for the former ($str$ meaning 
``structural'') and $t^{2}_{obs}$ for the 
latter. Photoionization models of planetary nebulae and \hii\ regions generally
fail to produce such large values of $t^{2}_{obs}$ as observed in 
real nebulae (e.g. Kingdon \& Ferland 1995, P\'{e}rez 1997),
except in the case of high metallicities, i.e. equal 
to the canonical ``solar'' value or larger. Note that in this case, 
what produces $t^{2}_{obs}$ in the model is actually the temperature gradient 
discussed above. 
Density fluctuations  could be a source of temperature fluctuations, 
due to increased collisional deexcitation in zones of higher density,
 but photoionization models 
including such density fluctuations  also fail to return large enough values of 
$t^{2}$ (Kingdon \& Ferland 1995). Note that introducing a density 
condensation shifts the dominant oxygen ion to a less charged one, 
and consequently the increase in $t^{2}(\rm O^{++})$ is not as 
important as might have 
been thought a priori.  Viegas \& 
Clegg (1994) argued that very high density clumps ($n$  $>$ 
$10^{5}$\cmcub) 
could \emph{mimic} the effects of temperature fluctuations by collisionally 
deexciting the \Oiii\ line.  The existence of such clumps is however 
not confirmed by the densities derived from \rAriv, from the ratio of 
fine structure \oiii\ lines and from high order Balmer decrement lines (Liu et 
al. 2000, 2001).  Note that, if they existed, such clumps should be located very 
close to the star in order to emit significantly in \Oiii\ with respect 
to the rest of the nebula. 

As will be discussed in Sect. 3.6, abundance inhomogeneities have been proposed to solve the ORL / CEL 
problem (Torres-Peimbert et al. 1990, P\'{e}quignot et al. 2002).  
Carbon and/or oxygen rich pockets would produce zones of lower 
temperature (due to increased cooling).  In PNe the existence of carbon rich 
pockets is attested from direct observations in at least a few objects
(e.g. Abell 30 and Abell 78, Jacoby \& Ford 1983) and these carbon rich 
inclusions are thought to be material coming from the third dredge up 
in the progenitor star (Iben et al. 1983). But the existence of oxygen rich pockets
 in PNe is more difficult to understand
from the present day evolution models for intermediate mass stars (see Sect. 5).  On the other hand, in \hii\ regions, especially in 
giant \hii\ regions, oxygen rich pockets could be made of material ejected by 
Type II supernovae and not yet mixed with the gas (Elmegreen 1998).

Other origins of temperature fluctuations have also been proposed, 
involving additional heating processes.  The fact that several 
detailed photoionization studies of planetary nebulae (Pe\~{na} et al. 1998) or 
giant \hii\ regions 
(Garc\'{\i}a-Vargas et al. 1997, Stasi\'{n}ska \& Schaerer 1999, 
Luridiana, et al. 1999, Luridiana \& Peimbert 
2001, Rela\~{n}o et al. 2002) predict significantly lower \rOiii\
ratios than observed indeed argues for additional heating.  
Shock heating or conductive heating are among the possibilities to 
investigate. Heat conduction from hot bubbles  has been examined by 
Maciejewski et al. (1996) and shown to be
insufficient to explain the   $t^{2}$ derived from observations.
The 
energy requirements to produce the observed values of $t^{2}$ have 
been evaluated by Binette et al. (2001) in the hypothesis of hot spots 
caused by an unknown heating process. 
In \hii\ regions, the mechanical energy 
associated with the ionizing sources (stellar winds, supernova 
explosions) does not 
seem sufficient to produce the required value of $t^{2}$ (Binette et 
al. 2001, Luridiana et al. 2001).  
In planetary nebulae, a considerable amount of kinetic energy is 
available from the central star winds. The radiative hydrodynamical models 
of Perinotto et al. (1998) present
 a temperature spike at the external shock front. This temperature 
 increase, located in a relatively narrow external zone,
  is not expected to produce a higher 
 \rOiii\ temperature than derived from the Balmer jump. In the radiative 
 hydrodynamical models computed by Mellema \& Frank (1995) for 
 aspherical nebulae, there are areas of lower  density in which
  cooling is inefficient and the temperature is 
higher due to shock heating. Mellema \& Frank  suggest that this may explain the 
differences in temperatures derived from different methods. However, 
a quantitative analysis remains to be done in order to check whether 
the predicted effect indeed reproduces what is observed. Simulations 
taking into account the evolution of the velocity and mass-loss rate 
of the fast central star wind (Dwarkadas \& Balick 1998) lead to 
considerably more structure on smaller scales, which could be even 
more favorable to solve the temperature fluctuation problem. 
In a slightly different context, Hyung et al. (2001) have tried to 
explain the high temperature observed in the inner halo of NGC 6543, 
 (15\,000~K as 
opposed to ~8\,500~K for the bright core) by means of a  simulation 
using a 
hydrodynamic code coupled to a photoionization calculation. 
These authors showed that mass loss and velocity 
variations in the AGB wind can simultaneously explain the existence of shells 
in the halo and 
the higher \Opp\ temperature.

Recently, Stasi\'{n}ska \& Szczerba (2001) proposed a completely 
different origin for temperature fluctuations,
 related to photoelectric heating by dust grains. This hypothesis is 
 also very promising and can be checked observationally (see Sect 3.7.5).

Although the   $t^{2}$ scheme has proved very useful to uncover the 
possibile existence of
temperature inhomogeneities, it may not be appropriate 
to describe the real situation.  In the case where abundance 
inhomogeneities are  the source of the temperature variations the Peimbert (1967) 
description is obviously inadequate. But it can also be inappropriate
for nebulae of homogeneous chemical composition, as shown 
on a simple two-component toy model. 
Consider two homogeneous zones of volumes $V_{1}$ and $V_{2}$ with 
temperatures $T_{1}$ and $T_{2}$, electron densities $n_{1}$ and 
$n_{2}$, and densities of the emitting ions (e.g.  \Opp) $N_{1}$ and 
$N_{2}$.  Calling $f$ the ratio $(N_{2} n_{2} V_{2})/(N_{1} n_{1} 
V_{1})$ of the weigths of the emitting regions, the mean electron 
temperature defined by Peimbert (1967) can be expressed as: 
\begin {equation}
 T_{0}= \frac{ T_{1} + 
f T_{2}} {1+f} 
\end{equation}
and $t^{2}$ as:
\begin {equation} t^{2}= \frac{ 
(T_{1}-T_{0})^2 + f (T_{2}-T_{0})^2} {(1+f)T_{0}^2}.
\end {equation}
For $T_{0}$=10\,000~K, the case $f=1$ 
(i.e. regions of equal weight) corresponds to $T_{1}$=12\,000~K and 
$T_{2}=$8\,000~K. It must be realized that 
this 4\,000~K difference in temperatures requires a 
difference of a factor 3 in the heating or cooling rates between the 
two zones. When $f \gg 1$, there is a high weight zone at 
$T_{2} \le T_{0}$ and a low weight zone at $T_{1} \gg T_{0}$.  Such a 
situation could correspond to a photoionized nebula with small volumes 
being heated by shocks or conduction.  When $f \ll 1$, there is a high 
weight zone at $T_{1} \ge T_{0}$ and a low weight zone at $T_{2} \ll 
T_{0}$ which could correspond to high metallicity clumps.
With such a toy model, one can explore the biases in abundance 
obtained for \Opp\ using the  \rOiii\ temperature and different 
lines emitted by this ion. Examples 
are shown in Figs. 8 and 9 of Stasi\'{n}ska (2002). 
 Following expectations, \Opp\ derived from \Oiii\ is generally 
underestimated, but it is interesting to note that the magnitude of 
the effect depends both on $f$ and on $T_{0}$.  The bias is 
very small when $T_{0}$ $\gtrsim$ 15\,000~K. It is small in any case if 
$f > $ 3 -- 4, because \Oiiit\ saturates above $\sim$ 50\,000K. At 
$T_{0}$ $\sim$  8\,000~K, \Opp\ is underestimated by  
 up to a factor of 2--3 in the regime where \Oiiit\ is significantly 
 emitted in both zones.  
As expected,  \Opp\ derived from infrared 
fine structure lines  and from the optical recombination line 
{O~{\sc ii} $\lambda$4651} is correct.
Such a toy model demonstrates that the classical temperature 
fluctuation scheme can be misleading.  Even in a simple two zone 
model, the situation needs at least three parameteres to be described, not two.  In 
our representation, these parameters would be $T_{1}$, $T_{2}$ and 
$f$, but other definitions can be used.    

 \subsection{The  optical recombination lines mystery}
 
 It has been known for several decades that 
optical recombination lines in PNe and \hii\ regions indicate higher 
abundances than collisionally 
excited lines (see Liu 2002 for a  review).  Most of the 
former studies concerned the carbon 
abundance as derived from C~{\sc ii} $\lambda$4267 and from \Ciii, but more recent 
studies show that the same problem occurs with lines from \Opp, \Npp\ and 
\Nepp\ (Liu et al. 1995b, 2000, 2001, Luo et al. 2001). 
The ORL abundances are 
higher than CEL abundances by factors of about 2 for most PNe, 
discrepancies over a factor 5 are found in about 5~\% of the PNe 
and  can reach factors as large as 20 (Liu 2002). For a given 
nebula, the discrepancies for the individual elements C, N, O, Ne are 
found to be approximately of the same magnitude.

The explanations most often invoked are: i) temperature 
fluctuations, ii) incorrect atomic data , iii) fluorescent excitation, iv) 
upward bias in the measurement of weak line intensities, v) blending with other lines, 
vi) abundance inhomogeneities. None of them is completely 
satisfactory, some are now definitely abandoned.

The completion of the OPACITY project has allowed accurate computation of 
 effective recombination coefficients needed to analyze ORL data.
 The advent of high quantum efficiency, large dynamic range and 
large format CCDs now allows to obtain high quality 
measurements of many faint recombination lines for bright PNe, thus 
hypothesis iv) cannot be invoked anymore. In addition, 
Mathis \& Liu (1999) have measured  the 
weak [O~{\sc iii}] $\lambda$4931, whose intensity ratio with \Oiii\ 
depends only on the ratio 
of transition probabilities from the upper levels. They found  
$(4.15 \pm 0.11)~10^{-4}$ compared to the theoretical values 
$4.09~10^{-4}$ (Nussbaumer \& Storey 1981), $4.15~10^{-4}$
(Froese Fisher \& Saha 1985), 
$2.5~ 10^{-4}$ (Galav\'{\i}s et al. 1997).  If, as expected, the latter 
computations give the more accurate results, the bias in the 
measurement of extremely weak lines could amount to 60\%. This is far 
below what is needed to explain the ORL/CEL discrepancy.  
A large number of faint 
recombination lines have now been measured, and the observed relative 
intensities of permitted transitions from \Cpp, 
\Npp, \Opp\ and \Nepp\ are in 
agreement with the predictions of recombination theory, which goes 
against ii), iii), iv) and v).   As mentioned in the previous 
subsection, 
the  values of $t^{2}$ derived from the comparison of temperatures from \rOiii\ 
and from the Balmer jump are too small to account for the large 
abundances derived from the ORL, therefore i) does not seem to be the 
good explanation. This is true even adopting a two-zone toy model 
instead of  Peimbert's fluctuation scheme.

On the basis of detailed studies of several PNe, Liu (2002) 
notes that for a given 
nebula, the discrepancies for the individual elements C, N, O, Ne are 
found to be approximately of the same magnitude.
The ORL/CEL abundance ratios  
correlate with the difference between the temperatures from \rOiii\ 
and from the Balmer jump. Liu et al. 
(2000) favour the hypothesis of an inhomogeneous composition, with 
clumps having He/H = 0.4 and C, N, O, Ne abundances around 400 times 
that in the diffuse gas in the case of NGC 6153.  It
 is indeed possible to construct a 
photoionization model with components of different 
chemical composition that reproduces 
the observed integrated line ratios satisfactorily (P\'{e}quignot et al. 2002).
  However, such a model is 
difficult to reconcile with the present theories of element production 
in intermediate mass stars (e.g. Forestini \& Charbonnel 1997).  Also, such 
super metal rich knots are not in pressure equilibrium with the 
surroundings and should be short lived, unless they are very dense.

Spatial analyses of NGC 6153 (Liu et al. 2000) and of NGC 6720 (Garnett 
\& Dinerstein 2001) show that the ORL/CEL discrepancy 
decreases with distance to the central star.
A possible explanation for the large intensities of the recombination 
lines of C, N, O, Ne, mentioned by Liu et al. (2000), is high 
temperature dielectronic recombination for states with high quantum 
numbers, a process so far not included in the computations of 
the effective recombination 
coefficients.  Then, the ORL would be preferentially emitted 
in regions of temperatures of (2 -- 5)~$10^{4}$~K.  There 
remains to find a way to obtain such high temperature material in 
planetary nebulae. Apart from conduction and shock fronts, 
there is also the possibility of heating by dust grains (see Sect. 
3.7.7).

 \subsection{The role of internal dust}
 
Until now we have omitted the solid component of nebulae, which, although 
not important by mass (usually of the order of $10^{-3}$, see Hoare et 
al. 1991, Natta \& Panagia 1981, Stasi\'{n}ska \& Szczerba 1999)
importantly affects the properties of PNe and \hii\ regions.  
The discussion below  deals only with  aspects that 
are explicitly linked with the derivation of the chemical composition of nebulae.

	\subsubsection{Evidence for the presence of dust in the ionized 
	regions}
	
Numerous mid- and far- IR spectral observations of PNe and \hii\ regions 
have shown the presence of a strong continuous emission at a 
temperature around 100 -- 200~K, attributed to dust grains 
heated by the ionizing stars (see the discussion in Pottasch 1984).  
Near- and mid-IR array observations have shown that the 
distribution of this emission is comparable to the distribution 
of [Ne\,{\sc{ii}}]\,$\lambda$12.8\,$\mu$m and 
[S\,{\sc{iv}}]\,$\lambda$10.5\,$\mu$m
 radiation, implying that dust is not only found in 
the neutral outskirts, but also inside the ionized regions (see review 
by Barlow 1993).  This does not necessarily imply, 
however, that  grains are intimately mixed with ionized gas.  A priori, 
they could be located exclusively in tiny, dusty neutral clumps, such as observed in the 
Helix nebula NGC 7293 (e. g. O'Dell \& Handron 1996) or in the Ring Nebula NGC 6720  
(Garnett \& Dinerstein 2001).  A crucial piece of evidence is 
provided by the following argument. It has been demonstrated by Kingdon et al. 
(1995) and  Kingdon \& Ferland (1997) that, in nebulae of normal 
chemical composition, numerous lines of elements such as Mg, Al, Ca, 
Cr, Fe,  should be detectable in ultraviolet or optical spectra.  What 
observations show is that these elements are depleted 
in PNe by factors around 10 -- 100 (Shields 1978, Shields et al. 
1981, Shields 1983, Harrington \& Marionni 1981,
Volk et al.  1997, Perinotto et al. 1999, Casassus et al.  2000).  
The same holds for \hii\ regions (Osterbrock et al. 1992, Esteban et al.  
1998).  

	\subsubsection{Heavy element depletion}
	
One important consequence of the above mentioned observational fact is 
that  analyses of ionized nebulae do not provide the real 
abundance of such elements as 
 Mg, Al, Ca, Cr, Fe, which can be incorporated in grains.  Carbon can also be 
significantly depleted in carbon-rich grains -- graphite or PAHs.  The 
measurement of carbon abundances from nebular lines therefore 
 gives only a lower limit to the total carbon content.  This is also 
true for oxygen, although to a much smaller extent.  In \hii\ regions it 
is possible to estimate the amount of oxygen trapped in dust grains 
from the observation of the Mg, Si and Fe depletions (see  Esteban et 
al.  1998).  Also, the consideration of the Ne/O ratio can be useful, 
since Ne, being a noble gas, cannot enter in the composition of grains.

	\subsubsection{The effect of dust on the ionization structure}

Dust internal to \hii\ regions and PNe competes with the gas in absorbing 
 Lyman continuum photons, therefore lowering the \Hb\ luminosity. 
The nebular ionization structure is affected by two competing 
processes. The ionization 
parameter drops due to the fact that part of the Lyman continuum photons 
are absorbed by dust and not by gas.  This alone 
would tend to lower the general ionization level.
The ionizing radiation field seen by the atomic 
species depends on the wavelength dependence of the dust absorption 
cross section.  For conventional dust properties, the absorption cross 
section per H nucleon smootly decreases for energies above 13.6~eV 
(see e.g. Fig. 1 from Aanestad 1989), favouring the ionization of He 
with respect to H. In the model of the Orion nebula
presented by Baldwin et al.  (1991), the net effect of absorption by 
dust is to bring the \Hp\ and \Hep\ zones into closer agreement.

	\subsubsection{The effect of dust obscuration on the emission line 
	spectrum}
	
The presence of dust inside the ionized regions affects the emission 
line spectrum by selectively absorbing (and scattering) the emitted 
photons. Since the emission lines from various ions are formed in different 
zones, their relative fluxes as measured by the observer do not only 
depend on a general extinction law, but also on the differences in the 
geometrical paths of the photons in the different lines.  This, in 
principle, can be modelled using a photoionization code including
dust but the problem is complex and the solution extremely 
geometry-dependent. For practical purposes, as explained in 
Sect. 3.3, it is more convenient 
 to deredden an observed spectrum by 
adjusting the observed Balmer decrement to a theoretical one. If 
comparisons need to be made with a photoionization model, 
then they should be made with the theoretical emitted  spectrum 
without dust attenuation. Of course, such a procedure is only 
approximate.	
	
Resonance lines have an increased path length with respect to other lines, 
and are therefore subject to stronger absorption by dust.  This is 
the case of H Ly$\alpha$, which may be entirely trapped 
by grains in the case of very dusty nebulae (Ly$\alpha$ absorption is actually 
one of the main heating
agents of dust particles in planetary 
nebulae, see e.g. Pottasch 1984).  Other 
resonance lines, such as C~{\sc iv} $\lambda$1550, N~{\sc v} 
$\lambda$1240 or Si~{\sc iv} $\lambda$1400, are also affected by this 
selective absorption process. Usually, 
an escape probability formalism is used to account for it (Cohen et 
al. 1984).  The observed intensity of the resonance lines depends on the 
amount of dust, on the ionization structure and 
on the velocity fields both in 
the nebula and in the surrounding halo and intervening 
interstellar medium (see e.g. Middlemass 1988). The inclusion of 
dust attenuation in a tailored photoionization model of the  PN 
NGC 7662 results in a derived gas phase C abundance  twice as large as 
would be deduced using classical methods (Harrington et al.  1982).

Another consequence of selective dust absorption is that
it prevents the 100~\% conversion of
high-n Lyman lines into Ly$\alpha$ and Balmer lines (the case B).  For dusty 
environments such as the Orion Nebula, the \Hb\ emissivity can be 
reduced by 15~\% (Cota \& Ferland 1988).

	\subsubsection{The effects of grains on heating and cooling of the 
	gas}
	
An obvious effect of the presence of grains on the thermal balance of 
ionized nebulae, is due to the depletion of strong coolants such as Si, Mg, 
Fe, which enhances the electron temperature with 
respect to a dust-free situation.  This aspect is important not only for 
detailed model fitting of nebulae, but also when using grids of 
photoionization models to calibrate strong line methods for abundance 
determinations (Henry 1993, Shields \& Kennicutt 1995).

Grains have also a \emph {direct} influence on the energy balance. 
The photoelectric effect on dust grains has been shown by 
Spitzer (1948) to be a potential heating source in the interstellar 
matter.
Baldwin et al. (1991) have introduced the physical effects of dust in 
the photoionization code CLOUDY. They constructed a detailed model of 
the Orion nebula and found that in this object, 
 heating by photoelectric effect can amount 
to a significant proportion of the total heating while
 collisions of the gas particles with the 
grains contribute somewhat to the cooling.

The effect of dust heating is dramatic in the H-poor and extremely dusty 
planetary nebula  IRAS 18333-2357 in which $m_{\rm d}/m_{\rm H}$ 
is estimated around 0.4 (Borkowski \& Harrington 1991).  
In this object, heating is almost entirely due to  
photoelectric effect. 

In nebulae in which dust-to-gas mass ratio, dust properties and grain size 
distributions have the canonical values, the relative importance 
 of dust heating is generally not very 
large.
If, however, there is a large proportion of \emph {small} dust grains, 
then the contribution of dust heating to the total energy gains
 may become  important, 
 as demonstrated by Dopita \& Sutherland (2000) 
on a grid of dusty photoionization models of planetary nebulae. The 
effect is more pronounced in the central parts of the nebulae, being 
proportional to the mean intensity of the ultraviolet radiation 
field, and gives rise to a strong temperature gradient.

 If such small grains do exist (and there is now growing evidence for that 
(Weingartner \& Draine, 2001), their presence in planetary 
nebulae would solve a number of problems that have found no 
satisfactory solution so far (see Stasi\'{n}ska \& 
Szczerba 2001): i) the thermal energy deficit 
inferred in some objects from tailored photoionization modelling;
 ii) the large negative 
temperature gradients inferred directly from spatially resolved observations 
and indirectly from integrated spectra in some PNe; iii) the 
Balmer jump temperatures being systematically smaller than temperatures
derived from forbidden lines; iv) the intensities 
of [O\,{\sc{i}}]\,$\lambda$6300 often observed to be larger than predicted by 
photoionization models: indeed, near the ionization 
front, Lyman continuum photons are exhausted and the only 
photons still present are photons below the 
Lyman limit. Those are not 
absorbed by hydrogen but can heat the gas via photoelectric 
effect on dust grains. One should however remember that dust is not the only 
way to enhance 
[O\,{\sc{i}}] emission, other mechanisms have been mentioned in
Sect. 1.1.

The energy gains per unit 
volume of gas due to  photoelectric effect, $G_{\rm d}$, 
are proportional to the number density 
of dust grains and to the intensity of the stellar radiation field. 
Combining with Eq. (1.16) which expresses the gains due to
photoionization of hydrogen, $G_{\rm H}$, 
 it is easy to show that  $G_{\rm 
d}/G_{\rm H}$ is  proportional to $(m_{\rm d}/m_{\rm H}) U$, where $U$ 
is the ionization parameter.
This has important consequences 
in the case of filamentary structures. If small grains are present, 
the photoelectric effect will boost the electron temperature in the 
low density components. This will result in important small-scale temperature 
variations in the nebula. The models of Stasi\'{n}ska \& 
Szczerba (2001) show that \emph{moderate}
 density inhomogeneities (such as inferred from high resolution images 
of PNe) give rise  to values of $t^{2}$ similar 
to the ones obtained from observations.  Note that, 
contrary to the dust-free case, the tenuous component has a higher 
\Te\ than filaments or clumps, therefore the clumps are better 
confined. 

 Stasi\'{n}ska \& Szczerba (2001) also point out that if, as expected,  dielectronic recombinations for high level states
 strongly enhance the emissivities of 
recombination lines, the presence of small grains in 
filamentary planetary nebulae would boost the emission of recombination 
lines from the diffuse component, principally in the inner zone.
Therefore, small grains could solve
 in a natural way both the temperature fluctuation problem and the ORL/CEL 
discrepancy.

The presence of small grains in planetary nebulae can be tested observationally by
 measuring the temperature {\em in} and {\em 
between} filaments.

 \subsection{The specific case of the helium abundance determination }
 
 The determination of the helium abundance follows the same principles 
as that of other elements.  But one is much more demanding about 
the accuracy.
To follow the production of helium in stars, and the 
evolution of the helium content in galaxies, 10\% accuracy is a goal
that one would like to achieve.  Helium abundances compiled from the 
literature over the years must be considered with caution, because 
of the different treatments adopted by various authors. On the other 
hand, the required accuracy should be reachable with consistent observations and 
modern data treatments.
To determine the primordial helium abundance, Y$_{p}$, 
 one needs a much better accuracy, since quite different cosmologies are 
 predicted for values Y$_{p}$ differing by a only few percent. 
From low metallicity giant extragalactic \hii\ regions, Olive et al. 
(1997) find Y$_{p}$ = 0.234 $\pm$ 0.002 while Izotov \& Thuan (1998) 
find Y$_{p}$ = 0.244 $\pm$ 0.002.  These two estimates are mutually exclusive.  
Is it possible to say which of the two ­- if any ­- is correct?

The first step is to obtain the intrinsic values of the intensities 
of the helium  and hydrogen lines in an observed spectrum. If the 
spectrum contains stellar light, as in the case of giant  \hii\ 
regions, one must correct the observed intensities for 
underlying stellar absorption.  The recent evolutionary synthesis models of 
Gonzalez 
Delgado et al. (1999) provide a theoretical framework for 
that.  One also has to correct the intensity ratios for reddening, 
assuming a given reddening ``law'' and a given intrinsic value 
of the ratios of the hydrogen line intensities.  The latter mainly depends on the electron 
temperature, which can be estimated from the \rOiii\ ratio, with a 
correction due the fact that the \Opp\ region is only a part of the \Hp\  region.  Using an appropriate number of lines, one can 
estimate iteratively the reddening and the correction for underlying 
absorption (e.g.  Izotov \& Thuan 1998).  However, as commented by 
Davidson \& Kinman (1985) and Sasselov \& Goldwirth (1995), and as 
mentioned in Sect. 3.3,
collisional excitation of H Balmer lines may become important, 
especially in  \hii\ regions of high \Te.  So far, this effect has always been 
omitted in the determination of the abundance of primordial He.  It 
may induce an 
overestimation of the reddening, and therefore an underestimation of 
the \Hep\ abundance derived from \Hei\ by up to 5~\%
(Stasi\'{n}ska \& Izotov 2001). The importance of this effect depends 
on the abundance of residual \Ho.  

Then, from the corrected ratios of emission lines one has to determine the value of 
\Hep/\Hp, or, to be more exact, of
$\int_{}^{} n({\rm He^{+}})  {\rm d}V / \int_{}^{} n({\rm H^{+}}) {\rm d}V$.
This assumes that the line emissivities do not vary 
strongly over the nebular volume.  The emissivities depend on \Te, 
and also on $n_{e}$ in the case of some helium lines, 
due to enhancement by collisional 
excitation from the metastable 2$^{3}$S level.  If one knows $n_{e}$ from plasma 
diagnostics, the contribution of collisional excitation can be 
obtained.  The recent tables of Benjamin et al. (1999), 
 based on a resolution of the statistical 
equilibrium of the He atom using the best available atomic data,
 can be used for this purpose. Note that these authors also 
  provide analytical fits, with the warning that some of them lead to values that may 
  differ by  1\% or more 
from the actually computed values of the emissivities.  
 Some  He line emissivities are also affected by self absorption of 
the pseudo resonance lines from the 2$^{3}$S level.  Using a sufficient number of helium 
lines, one can in principle  determine iteratively and self-consistently the 
characteristic temperature and density of the helium line emission, 
and the relative abundance of \Hep.  The treatment of 
radiation transfer in the lines remains to be improved  and is 
announced as a next step by Benjamin et al. (1999).  However, this is 
a complex problem: it depends on the velocity field and on 
the amount of internal dust which selectively absorbs resonant 
photons.  Therefore, one does not expect models to be easily 
applicable to real objects.   
However, since this is a second order effect, this is perhaps not too 
problematic, if one discards the lines  likely to be most 
affected by this process.  One must not forget that the 
emissivities of the H Balmer lines too may be in question, both 
 because of the contribution of collisional excitation mentioned above 
 and because the presence of dust deviates the hydrogen spectrum from case 
B (see Hummer \& Storey 1992).
Another problem is to take into account the non uniformity of \Te.  
Sauer \& Jedamczik (2002) 
have computed a grid of photoionization models for this purpose, and 
introduce the concept of a ``temperature correction factor'' which 
they compute in their models.  Note, however, that the real 
temperature structure of nebulae 
 is not obtained 
from ``first principles'', as the preceding sections made clear.  
Therefore, the distribution of \Te\ in real objects has most probably a larger
impact than predicted by the models of Sauer \& Jedamczik (2002). 
Peimbert et al. (2002) have adopted a semi-empirical approach, based 
on the Peimbert's (1967) temperature fluctuation 
scheme. But the temperature fluctuation scheme may give spurious 
results in the hypothesis of zones of highly different temperatures, as 
argued in Sect. 3.5.2. 

If \heii\ lines are present in the spectra, they have to be accounted 
for, to determine  
$\int_{}^{} n({\rm He^{++}})  {\rm d}V / \int_{}^{} n({\rm H^{+}}) {\rm d}V$.
The major uncertainty in that case comes probably from the lack of 
knowledge of the temperature characterising the emission of \heii\ 
lines.  An additional difficulty is due to the fact that part of the 
\heii\ emission may be of stellar origin.

The He/H abundance is obtained after considering ionization 
structure effects.
For low values of the mean effective temperature of the radiation 
field, a zone of neutral helium is present. Unfortunately, no 
ionization correction formula can be safely applied, since 
 the ionization structure of helium with respect to hydrogen mainly 
 depends on the hardness of the radiation field, while the ionization 
 structure of the heavy elements also strongly depends on the 
gas distribution (e.g. Stasi\'{n}ska 1980b). In the 
case of an \hii\  region ionized by very hot stars, photoionization 
models show that the \Hep\ region may 
on the contrary extend 
further than the \Hp\ region (see for example
Stasi\'{n}ska 1980b or  Sauer \& Jedamczik 
2002). Whether this is the case for an object under study should be 
tested by models. 

Olive \& Skillman (2001) stress the importance of having a 
sufficient number of observational constraints and of using them 
in a self consistent manner  with a 
Monte-Carlo treatment of all sources of errors. Unfortunately, the 
 errors on the temperature structure and on the ionization 
structure of real nebulae are very difficult to evaluate, and this, 
combined with uncertainties in atomic parameters and deviations from 
case B theory implies that the uncertainty in derived helium 
abundances is certainly larger than claimed.

\section{Observational results on abundances in  H~{\sc ii} regions of the 
Milky Way}

 \subsection{The Orion nebula: a benchmark}
 
The Orion nebula is the brightest and most observed \hii\ region in the galaxy.  
Therefore it is a benchmark in many respects. O'Dell (2001) and 
Ferland (2001) have summarized our present knowledge on this object. 
In the following, we 
only discuss aspects related to the chemical composition in the 
ionized gas.  

It is of interest, beforehand, to mention that it is with the 
 Orion nebula that the concept of filling factor 
 started. Using a spherical representation, Osterbrock \& Flather 
 (1959) showed that the optical surface brightness data could be 
 reconciled with the observed \rOii\ intensity ratios only when 
 assuming extreme density fluctuations. They proposed a schematic 
 model in which these fluctuations are represented as condensations 
 immersed in a vacuum, with the relative volume of the condensations 
 being only 1/30 of the total volume of the nebula.
 But a more realistic model of the Orion nebula (Zuckerman 
 1973, Balick et al. 1974, see also discussion in O'Dell 2001) is to 
 represent the Orion nebula as 
 an ionized blister on a background molecular cloud.  
From a detailed comparison of the \Hb\ surface brightness map and of 
the \rSii\ map,  Wen \& O'Dell (1995) derived a 3D representation of the nebula.  
The ionized skin is very thin with respect to the overall size of the 
nebula, which justifies the plane parallel approximation for 
photoionization modelling.

The extinction in the Orion nebula is well known to differ from 
the standard reddening law, and  has been studied in detail 
(see  Baldwin et al. 1991, Bautista et al. 1995, Henney 1998 for 
recent references)

Abundances have been derived both from 
\Te-based empirical methods and from photoionization 
models, using optical data with increased signal to noise and spectral 
resolution, with the addition of ultraviolet data obtained with IUE (and more 
recently with HST) and infrared data from ground-based telescopes and 
from  KAO, ISO, and MSX. Table 2 summarizes the abundances derived during 
the last decade. All the abundances are given in ppM units 
($10^{6}$ $\times$ the number of 
particles of a given species with respect to hydrogen) 

There is rather good agreement for the oxygen abundances, the value of Esteban 
et al. (1998)  with $t^{2}$ = 0 being the lowest and the one with  $t^{2}$=0.024 being the 
highest. Note that the preferred abundances of Esteban et al. (1998) 
are those obtained with  $t^{2}$=0.024, which is the value indicated 
by the ORL/CEL comparison. However, the comparison of the \rOiii\   
and Balmer  jump temperatures is consistent with $t^{2}$ = 0.    
One must be aware that abundances from models are not always the most reliable, 
since the models do not reproduce the ionization structure perfectly. 
The values of  Simpson et al. (1998) for Ne, S, and Ar are obtained 
from simultaneous observations of the most abundant ionic stages.

The Mg, Si, Fe and Ni abundances are heavily depleted with 
respect to the Sun (indicating the presence of grains intimately mixed 
with the gas phase in the ionized region). 
There is actually a controversy with respect to the interpretation of 
Fe lines (Bautista et al. 1994, Baldwin et al. 1996, Bautista \& 
Pradhan 1998). Esteban et al. (1999) recommend to derive Fe 
abundances from Fe$^{++}$ lines as done in the works quoted in Table 2.

\begin{table*}
\begin{flushleft}
\begin{tabular}{lrcccccccccc} 
\hline

  & He & C & N & O & Ne & Mg & Si & S & Ar & Fe & Ni \\
	 & 		&		&		&		&		&		&		&		&		&		&		\\ 
a & 	100000	&	280	&	68 	&	400	&	81:	&		&	 4.5	&	8.5 	&	4.5	&		&		 \\ 
b & 	90000	&	210	&	87:	&	380	&	390:	&	3.2:	&		&	13.3	&	2.1	&	4.2:	&		 \\ 
c & 	101000	&		&	52:	&	310	&	40 	&		&	 	&	9.4	&	2.6	&	2.7:	&	0.14:	\\ 
d &       97700 & 250 & 60 & 440 & 78 & & & 14.8 & 6.3 & 1.3 &  \\
e & 	100000	&	250	&	42	&	300	&	50	&		&		&	9.3	&	3.3	&	2.2	&		\\ 
f & 		&		&		&		&	99	&		&	 	&	8.6	&	2.6	&		&		\\ 
g & 		&		&		&	326	&		&		&		&		&		&		&		 \\ 
\hline
\multicolumn{12}{l}{
\begin{minipage}{12cm}
\footnotesize {
a Rubin et al. (1991, 1993), optical + IR spectroscopy, model

b Baldwin et al. (1991), long slit optical + IR + UV spectroscopy, model

c Osterbrock et al. (1992), optical spectroscopy, empirical

d Esteban et al.  (1998) ($t^{2}$ = 0.024), optical spectroscopy, empirical

e Esteban et al.  (1998) ($t^{2}$ = 0.0), optical spectroscopy, empirical

f Simpson et al. (1998), IR spectroscopy, empirical

g Deharveng et al. (2000), optical integrated spectroscopy, empirical

}
 \end{minipage}}\\
\hline   
\end{tabular}
\end{flushleft}
\label{Table2}

\caption{recent measurements of the Orion nebula abundances (ppM units)}

\end{table*}



\begin{table*}
\begin{flushleft}
\begin{tabular}{lrcccccccccc}
\hline   
       
    & He & C & N & O & Ne & Mg & Si & S & Ar & Fe & Ni \\
     \multicolumn{12}{l}{Sun}  \\   
a & 98000: & 363 & 112 & 851 & 123: & 38 & 35 & 16 & 3.6: & 47 & 1.8  \\
b & 	85000:	&	331	&	83	&	676	&	120:	&	38	&	35	&	21	&	2.5:	&	32	&	1.8	 \\ 
c & 		&	391	&	85	&	544	&		&	34	&	34	&		&		&	28	& \\ 
d & 		&		&		&	490	&		&		&		&		&		&		&\\ 
     \multicolumn{12}{l}{Gas phase local interstellar medium} \\ 
e & 	 	&	141	&	75	&	319	&	 	&	22	&	19.5	&	16.6	&	 	&	7.4	&	0.26\\
     \multicolumn{12}{l}{Be stars}   \\
	 & 		&		&		&		&		&		&		&		&		&		&		\\ 
f	 & 		&	224	&	44.5	&	407	&		&	    	&	    	&		&		&	    	&		\\ 
g	 & 		&	190	&	64.7	&	350	&		&	23.0	&	18.8	&		&		&	28	&		\\ 
\hline
\multicolumn{12}{l}{
\begin{minipage}{12cm}
\footnotesize {
a Anders \& Grevesse (1989)

b Grevesse \& Sauval (1998)

c Holweger (2001)

d Allende Prieto et al. (2001)	

e Meyer et al. (1998) (O), Meyer et al. (1997) (N), Sofia et al. 
(1997) (C), 
Cardelli et al. (1996), Sembach \& Savage (1996)	(Si, S, Fe, Ni)	 
Sofia \& (Meyer 2001) (Mg)
	
f Cunha \& Lambert 1994

g Compilation from Sofia \& Meyer (2001)

}
 \end{minipage}}\\
 \hline 
\end{tabular}
\end{flushleft}
\label{Table3}
\caption{Solar vicinity abundances (ppM units) }
\end{table*}

 \subsection{Abundance patterns in the solar vicinity and the solar 
 abundance discrepancy}

Stars and nebulae provide a different perspective of the solar 
vicinity chemical composition.
The methods for abundance determinations differ (and might be in error 
in different ways) and the astrophysical significance of the abundances is not 
necessarily the same. 
One expects a priori 
the surface composition of the Sun to be identical with that of 
other objects in the 
solar vicinity. It turns out that the abundances from nearby \hii\ regions 
(Orion being the best example) are significantly 
smaller than the solar abundances 
from the works of Anders \& Grevesse (1989) or Grevesse \& Sauval 
(1998). It is to be noted that, despite of this fact, 
the reference abundance is often chosen to be that of the Sun 
from  Anders \& Grevesse (1989).  Table 3 
summarizes the abundances in the Sun, in the local interstellar 
medium (ISM) and in local B stars from recent references.

 Peimbert et al. (2001) notes that a decade ago, the oxygen abundance in 
the Sun was 0.44~dex higher than in Orion but when using the value from 
Esteban et al. (1998) with $t^{2}$ = 0.024 and the solar value of 
Grevesse \& Sauval (1998), the difference is only 0.19~dex.  When accounting for the 
fraction of oxygen that is contained in dust grains (which can be done 
assuming a standard chemical composition of the dust grains, and the 
constraints provided by the Mg, Si and Fe abundances), the oxygen 
abundance is multiplied by a factor 1.2 and the difference 
between the Solar value and Orion is only 0.11~dex.

The oxygen abundance in Orion obtained  with $t^{2}$ = 0 is actually similar 
to the one in the local 
interstellar medium (obtained from high resolution and 
high signal to noise absorption measurements, Meyer et al. 1998) and in local B stars
(e.g. Cunha \& Lambert 1994). Several possible explanations have been 
invoked. The ones listed 
by Meyer et al. 1998 are: i) an early enrichment  of the Solar system by a local 
supernova (not really tenable if the abundances of \emph {all} the elements in 
the local ISM are 2/3 solar); ii) a recent infall of metal poor gas in 
the local Milky Way;  iii) an outward diffusion of the Sun from a smaller 
Galactocentric distance.
A more recent discussion Sofia \& Meyer (2001) definitely rejects 
the hypothesis of the local ISM standard being 2/3 of the Sun.
  Indeed, new determinations give much smaller values for O/H: 
544~ppM  (Holweger 2001), 490~ppM (Allende Prieto et al. 2001).  The 
support for the 2/3 solar value is also invalidated from carbon (see their 
discussion). Note that Sofia \& Meyer (2001) also argue that B stars 
have metal abundances 
that are \emph {too low} to be considered valid representations of the ISM. 
According to Meyer et al. (1998), the local standard oxygen abundance 
 should be 540 ppM (gas + dust). 

In conclusion, the ``solar abundance discrepancy'' has gradually disappeared, 
mostly because modern derivations of the solar oxygen abundance give 
much lower values than earlier ones.   This reinforces confidence in
\hii\ regions as probes of the ISM abundances and in the methods 
used to analyze them. This is good news, since \hii\ regions are 
practically the only way to derive oxygen abundance in external 
galaxies, if one excepts the abundance analysis in giant 
stars of local galaxies which require very large telescopes. Giant \hii\ 
regions can be used as abundance indicators
up to large redshifts (see Pettini in the same 
volume).

 \subsection{Abundance gradients in the Galaxy from \hii\ regions}
 
 Abundance gradients in disk galaxies constitute one of the more 
important observational constraints for models of galaxy chemical 
evolution.  As a matter of fact, abundance gradients 
were first recognized to exist in external galaxies, where radial 
trends of emission line ratios were noted as far back as in the 
fourties 
(Aller 1942) and were attributed to abundance gradients in the 
early seventies (Searle 1971, Shields 1974).

In our own galaxy, gradients are more difficult to determine, due to
distance uncertainties and because many \hii\ regions are highly 
obscured by dust lying close to the galactic plane.  
The first determination of an 
abundance gradient in our galaxy from \hii\ regions was made by 
Peimbert, Torres-Peimbert \& Rayo (1978).
It is worth the effort to derive abundance gradients in the Milky Way 
because it is a benchline for chemical evolution of 
galaxies.  Only in the Milky Way can one have direct access to 
abundance measurements from so many sources as \hii\ regions, planetary 
nebulae, individual B, F, G stars etc..., which all probe different 
epochs in the Milky Way history.  Esteban \& Peimbert (1995) and Hou et al. 
(2000) provide excellent reviews on this topic. Table 4 presents a compilation of Galactic abundance gradients 
from H~{\sc ii} regions in units of d log(X/H) / dR in kpc$^{-1}$. 
Column 9 indicates the spanned range of galactocentric  
distances in kpc. Column 10 lists the total number of 
objects used to derive the gradients. Note that the errors quoted for  the gradients 
include only the scatter in the nominal values of the derived 
abundances about  the best fit line.  They do not take into account the 
uncertainties in the abundances and the possible errors on the 
galactocentric distances. Most abundances were obtained using 
empirical methods. 

\begin{table*}
\begin{flushleft}
\begin{tabular}[h]{lrrrrrrrrr}
\hline               
  & He & C & N & O & Ne & S & Ar & range & nb  \\
	 & 		&		&		&		&		&		&		&		&	     \\ 
a & 	0.02	&		&	-0.23	&	-0.13	&		&		&		&	8--14	&	5		\\ 
 	 & 	$\pm$~0.01	&		&	$\pm$~0.06	&	$\pm$~0.04	&		&		&		&		&		 \\ 
     & 		&		&		&		&		&		&		&		&		\\ 
b & 	-0.001	&		&	-0.090	&	-0.070	&		&	-0.010	&	-0.060	&	4--14	&	35\\ 
	 & 	$\pm$~0.008	&		&	$\pm$~0.015	&	$\pm$~0.015	&		&	$\pm$~0.020	&	$\pm$~0.015	&	& \\ 
     & 		&		&		&		&		&		&		&		&		\\ 
c	 & 		&		&		&		&	-0.086	&	-0.051	&		&	0-12	&	95	 \\ 
 	 & 		&		&		&		&	$\pm$~0.013	&	$\pm$~0.013	&		&		&	 \\ 
     & 		&		&		&		&		&		&		&		&		\\ 
d	 & 		&		&	-0.100	&		&	-0.080	&	0.070	&		&	0-10	&	23	\\ 
 	 & 		&		&	$\pm$~0.020	&		&	$\pm$~0.020	&	$\pm$~0.020	&		&		&		\\ 
     & 		&		&		&		&		&		&		&		&		\\ 
e	 & 		&		&	+0.002	&		&	-0.051	&	-0.013	&		&	12-18	&	15	 \\ 
 	 & 		&		&	0.020	&		&		&	0.020	&		&		&			 \\ 
     & 		&		&		&		&		&		&		&		&		\\ 
f	 & 	 	&		&		&	-0.047	&		&		&		&	0--17	&	28		\\ 
& & & & $\pm$~0.009 & & & & &  \\
     & 		&		&		&		&		&		&		&		&		\\ 
g	 & 		&		&	-0.072	&	-0.064	&		&	-0.063	&		&	0--12	&	34	\\ 
 	 & 		&		&	$\pm$~0.006	&	$\pm$~0.009	&		&	$\pm$~0.006	&		&		&			 \\ 
     & 		&		&		&		&		&		&		&		&		\\ 
h	 & 		&		&	-0.111	&		&		&	-0.079	&		&	0--17	&	28	 \\ 
 	 & 		&		&	$\pm$~0.012	&		&		&	$\pm$~0.009	&		&		&			 \\ 
     & 		&		&		&		&		&		&		&		&		\\ 
i	 & 	-0.004	&	-0.133	&	-0.048	&	-0.049	&	-0.045	&	-0.055	&	-0.044	&	6-9	&	3		 \\ 
 	 & 	$\pm$~0.005	&	$\pm$~0.002	&	$\pm$~0.017	&	$\pm$~0.017	&	$\pm$~0.017	&	$\pm$~0.017	&	$\pm$~0.030	&
 	 		&			  \\ 
     & 		&		&		&		&		&		&		&		&		\\  	 		
j	 & 		&		&		&	-0.040	&		&		&		&	5--15	&	34		 \\ 
 	 & 		&		&		&	$\pm$~0.005	&		&		&		&		&			 	 \\ 
     & 		&		&		&		&		&		&		&		&		\\ 
k	 & 		&		&		&		&	-0.039	&		&	-0.045	&	0--15	&	34				 \\ 
 	 & 		&		&		&		&	$\pm$~0.007	&		&	$\pm$~0.011	&		&				 \\ 
\hline
\multicolumn{10}{l}{
\begin{minipage}{12cm}
\footnotesize {
a Peimbert et al. (1978), optical spectroscopy, $t^{2}$=.035 

b Shaver et al.  (1983), 	optical spectroscopy for 
30 objects, radio data for 67 objects, $t^{2}$=0	

c Simpson \& Rubin  (1990), 	FIR  data from IRAS,	no icfs	

d Simpson  et al. (1995), 	FIR data from KAO, 	models  	

e Vilchez \& Esteban  (1996), 	long slit optical spectroscopy,   $t^{2}$=0

f Afflerbach  et al.  (1996), models to reproduce the $T_{e}$ measured 
from radio recombination lines in 28 ultracompact H~{\sc ii} regions		

g Afflerbach  et al. (1997), 	FIR data from KAO: 15 objects + sources from Simpson, models		

h Rudolph et al.  (1997), FIR data from KAO of 5 H~{\sc ii} regions in the outer 
Galaxy + results from Simpson	models			

i Esteban et al.  (1999), 	optical  echelle spectroscopy, $t^{2}$ $>$ 0	

j Deharveng et al.  (2000), absolute integrated optical fluxes,  $t^{2}$=0, rediscussion 
of distances 		 	

k Mart\'{\i}n-Hern\'{a}ndez et al.  (2002), FIR data from ISO, model grids, rediscussion 
of distances 
}
 \end{minipage}}\\
 \hline 
\end{tabular}
\end{flushleft}
\label{Table4}
\caption{Galactic abundance gradients from H~{\sc ii} regions d log(X/H) / dR in kpc$^{-1}$}
\end{table*}

It must be noted that, even in the case of similar methods, some details in 
the procedures employed may lead to significantly different results. 
For example, the much larger oxygen gradient found by Peimbert et 
al. (1978) probably results from their using the temperature 
fluctuation scheme (with  $t^{2}$ = .035). 

A possible flattening of abundance gradients in the outer disk has been 
mentioned by Fich \& Silkey (1991) and Vilchez \& Esteban (1996) but 
Rudolph et al. (1997) and 
Deharveng et al. (2000) find no clear evidence for that. 

The situation with the N/O ratio is not clear. N/O ratios determined from  
\Npp/\Opp\ using far infrared (FIR) lines (Simpson et al. 1995, Afflerbach et al. 
1997, see also Lester et al. 1987 and Rubin et al. 1988) are up to 
twice 
the values derived from \Np/\Op\ using optical data.  Actually, what is 
found is that  \Npp/\Opp\ is 
larger than \Np/\Op, so it cannot be an ionization correction factor 
problem.  Rubin et al. (1988) 
suggest that the discrepancy may be due to the neglect of the recombination 
component of the \Oii\ emission.  Such an explanation can indeed hold 
 at low \Te\ (say below 6\,000~K) but is  not expected to 
work at high \Te.  Another possibility suggested by Rubin et al. 
(1988) is 
that the  \Oiiitonea\ and \Oiiitoneb\ lines are optically thick, thus increasing 
the derived \Npp/\Opp.
FIR lines from \Npp\ and \Opp\ have now been observed by ISO (Peeters 
et al. 2002), but in their analysis Martin-Hern\'{a}ndez et al. (2002) do not use them 
to derive abundance gradients. It is not clear why, since they have 
constructed photoionization model grids to correct for unseen ions.

The only data on a possible carbon abundance gradient comes from 
optical recombination lines measures in 3 objects! Obviously more 
work is needed in this respect.

\subsection{The Galactic center}

The central parsec of the galaxy, identified with the Sagittarius A 
nebula, contains ionized gas powered by about $10^{40}$ ionizing 
photons~sec$^{-1}$ (Lacy el al. 1980).  A cluster of \hei\ emission line stars 
has been observed and spectroscopically analyzed (Tamblyn et al. 1996, 
Najarro et al. 1997).  The complete spectrum of 
infrared fine structure lines that has been observed, combined with 
the H Br$\alpha$ and Br$\gamma$ lines (see Shields \& Ferland 1994 for a 
compilation) should in principle allow to perform an abundance 
analysis.  From a two-component 
photoionization model Shields \& Ferland (1994)
estimate that the abundance of Ar 
should be about twice solar, but Ne seems rather to have the solar 
value.  The evidence for over solar metallicity is thus mixed. The  N/O 
ratio is estimated to about 3 -- 4 times solar.  However, the
 derived abundances may be clouded by errors in the 
reddening corrections (the extinction is as high as $A_{V}$=31, so, even at 
far infrared wavelengths, reddening  become important) and 
uncertainties in the atomic parameters (mainly those determining the 
ionization structure).  As a consistency check, Shields \& Ferland (1994) 
compared the electron temperature
measured from recombination lines with their model predictions.  For 
that, they included heating by dust, and assumed the same grain 
content as in the model of Baldwin et al. (1991) for Orion.  They found 
the measured temperatures to be consistent with a metallicity 1 -- 2 times solar, 
while 3 times solar would 
be only marginally consistent.  However, with a population of small 
grains, photoelectric heating would be more important, and larger metal 
abundances could be acceptable.

The Galactic center has since then been reobserved by ISO (Lutz et al. 
1996), but a detailed discussion of the new results remains to be 
done.
 
 \subsection{Nebulae around evolved massive stars}
 
 Evolved massive stars are associated with nebulae which result from 
the interaction of stellar winds and stellar ejecta with the ambient 
interstellar medium.  By studying the chemical composition of these 
nebulae, together with their morphology, kinematics and total gas 
content, one can get insight into the previous  evolutionary 
stages of the  stars and unveil some of the nucleosynthesis and mixing processes
occuring in their interiors.

Schematically, during main sequence evolution, the fast wind 
creates a cavity in the interstellar medium and sweeps out a shell of 
compressed gas.  After departure from the main sequence, the nature 
of the mass loss changes and the star loses chemically enriched 
material.  When 
the star reaches the Wolf-Rayet phase, its outer layers are 
almost hydrogen free. This material 
is lost at high velocity and catches up 
with material lost in previous stages (see Chu 1991  or 
Marston 1999 for a review).

Imaging surveys of the environments of WR stars have found that in 50\% 
of cases a ring like nebula is seen (Marson 1999).  Ring nebulae have 
been classified by Chu (1981) into R type -- radiatively excited \hii\ regions 
and subsonic expansion velocities, E type -- nebulae formed out of stellar 
ejecta (chaotic internal motion, large velocities) and W type  -- wind-blown 
bubbles showing thin sheets or filaments.  Atlases are published by 
Chu, Treffers \& Kwitter (1983), Miller \& Chu (1993) and Marston (1997).  
Known examples of R types are RCW 78 (amorphous, 
containing a WN 8 star) and RCW 118 (shell, surrounding a WN 6 star).  
Known cases of nebulae containing ejecta are M 1-67 (WN 8 star), 
RCW 58 (WN 8 star).  Known W types are NGC 6888 (WN 6 star), S 
308 (WN5 star), RCW 104 (WN4 star), although Esteban et al. (1992) 
consider NGC 6888 as an Bubble/Ejecta type in their classification.

Luminous Blue Variable stars are regarded as precursors of WR stars 
with the most massive progenitors.  They are usually found to be 
associated with small ejecta type nebulae like  $\eta$ Car, AG Car (Nota et 
al. 1995).

The first spatially resolved and comprehensive study of abundances in 
Wolf-Rayet ring nebulae is that of Esteban and coworkers  (Esteban et 
al. 1990, 1991, 1992, 
1993, Esteban \& V\'{\i}lchez 1992),
in which 11 objects have been analyzed with similar procedures.
In a plot relating the  N/O and O/H 
differential abundances (i.e. abundances with respect to  
interstellar medium ones)  Esteban et al. (1992) find that  
most objects lie close to the (O/H + N/H ) = (O/H + N/H)$_{\rm Orion}$ line, 
indicating  
that oxygen has been converted into nitrogen.  This is indeed what 
is predicted by the Maeder (1990) stellar evolution models at the 
beginning of the WN 
phase.
Dividing their objects into 3 categories from their chemical composition 
(\hii\ for objects with abundances close to those of the environing ISM, DN 
for diluted nebulae in which stellar ejecta are mixed with 
ambient gas and SE for pure stellar ejecta), Esteban et al. (1992) show 
that there is a rather good correspondence between the chemical 
classes and the morpho-kinematical classes.  They also 
note that the masses of SE nebulae are small and compatible with the 
hypothesis of pure stellar ejecta, while the \hii\ nebulae have 
larger dynamic ages, consistent with the idea of being composed 
of large quantities of swept up gas. Esteban et al. (1992) find that the SE nebulae 
surrounding WR stars are associated with stars showing variability and 
thus probably having unstable atmospheres.  This is also true for the 
nebulae associated with LBVs.
In plots relating the N/O mass fraction to the He mass fraction,
 Esteban et al. (1992) find that 
SE nebulae lie close to the  stellar 
evolution tracks of Maeder (1990) for initial masses 25 
-- 40\Ms, which become WN stars after a red supergiant (RSG) phase. 
This is consistent with the initial masses estimated from the star 
luminosities (Esteban et al. 1993).

Since this pioneering study, detailed computations have 
been performed to simulate the evolution of the circumstellar gas 
around massive stars (Garc\'{\i}a-Segura et al. 1996 a and b), 
coupling hydrodynamics with stellar evolution. The 
fate of the circumstellar gas results from interactions between the fast wind 
from the star while on the main sequence, the slow wind from the red supergiant or luminous blue variable 
stage and the fast wind from the WR stage. The resulting masses, 
morphologies and chemical composition of the circumstellar envelopes 
strongly depend on the initial 
stellar masses, both because of different nucleosynthesis and 
different time dependence of the winds. 
 Stars with initial masses 
around 35\Ms\ are predicted to go through a RSG stage, 
and produce massive nebular envelopes ($\sim$ 10\Ms) with composition only slightly 
enriched in He and CNO processed material. Stars with initial masses 
around 60\Ms\ are predicted to go through a LBV stage, 
and produce less massive  nebular envelopes  ($\sim$  4\Ms) with 
helium representing about 70\% of the total mass fraction, and 
CNO equilibrium abundances (C depleted by a factor 23, 
N enriched by a factor 13, and O depleted by a factor 18). The 
composition and morphology of NGC 6888 and Sh 308 well agree  with the 
theoretical prediction of a RSG progenitor. On the other hand, 
Smith (1996) notes that recent abundance determinations in 
 nebulae associated with LBV stars do
not agree with the composition predicted by the Garc\'{\i}a-Segura et al. (1996)
model of evolution of a 60\Ms\ star through the LBV stage. 
The abundance paterns of these nebulae are rather similar to those of 
SE nebulae surrounding RSG stars, with mild 
enrichments in He and N and mild depletion in O, suggesting that the 
star went through a RSG phase. It must be noted that abundance 
determinations in such objects are often difficult, because few diagnostic 
lines are available, so that ratios like N/H or O/H may be rather 
uncertain, but N/O is more reliable.
In a rediscussion of nebulae around LBV stars, Lamers et al. (2001) 
conclude that the stars have not gone through a RSG phase.  The 
chemical enhancements are due to rotation-induced mixing, and the 
ejection is possibly triggered by near-critical rotation.

\section{Observational results on abundances in planetary nebulae}

Until recently, a little less than 20 elements were available for 
abundance studies in planetary nebulae. These were: H, He, C, N, O, F, Ne, Na, 
Mg, Si, P, S, Cl, Ar, K, Ca, Mn, Fe, although routine abundance 
determinations are available for only about 10 elements. As already 
mentioned and will be made clearer in the next sections, these 
elements can serve either as probes of the ISM abundances or as probes 
of the nuclear and mixing processes in the progenitor stars. It has 
also been mentioned that some elements are heavily depleted in dust 
grains, so that the abundances of these elements in PNe (Mg, Si, P, K, 
Ca, Mn, Fe) rather give information on the chemistry of dust 
grains. This is of great interest since it is now believed 
that a large portion of grains found in the ISM were actually seeded 
in the atmospheres of evolved, intermediate mass stars (Dwek 1998). 

Recently, ultra deep spectroscopy of bright PNe allowed to 
detect and measure lines from elements of the fourth, fifth and even 
sixth row of 
the Mendeleev table (P\'{e}quignot \& Baluteau 1994, Baluteau et al.  
1995, Dinerstein 2001, Dinerstein \& Geballe 2001):
V, Cr, Co, Ni, Cu, Zn, Se, Br, Kr, Rb, 
Sr, Y, Te, I, Xe, Cs, Ba, Pb.  When the atomic data for a 
quantitative analysis of these lines become available (and some 
atomic physics work has already been fostered by these discoveries, 
see e.g. Sch\"{o}ning \& Butler 1998), this will open a new possibility to test PNe 
progenitors as production sites of r- and s- process elements. 

The determination of isotopic abundance ratios in PNe would allow 
serious constraints on the nucleosynthesis in post-AGB stars (see e.g. 
Forestini \& Charbonnel 1997, Marigo 2001). They strongly depend on 
stellar mass, metallicity and mixing length. Unfortunately, from the 
observational point of view, this field is still in its 
infancy. The $^{12}$C/$^{13}$C  ratio has been measured 
in only a couple 
of nebulae in either hyperfine UV transitions (Clegg et al. 1997, 
Brage et al. 1998) or in  millimetric lines of CO (Bachiller et al. 
1997, Palla et al. 2000). 
The $^{3}$He abundance has been determined in a few nebulae 
from the hyperfine transition at 
8.665~GHz (Balser et al. 1997, see also Galli et al. 1997) .

\begin{table*}
\begin{flushleft}
\begin{tabular}[h]{lrccccccccccccc}
\hline               

 	 & 	He	&	C	&	N	&	O	&	Ne	&	Na	&	Mg	&	Si	&	S	&	Cl	&	A	&	K	&	Ca	&	Fe	 \\ 
 	 & 	 	&	 	&	 	&	 	&	 	&	 	&	 	&	  	&	 	&		&		&		&		&        \\ 
a	 & 	10600	&	600	&	160	&	410	&	100	&	1.2	&	22	&	6.2	&	9.4	&	0.11	&	2.3	&	.05	&		&		 \\ 
b	 & 	11000	&	955	&	162	&	508	&	137	&		&		&		&		&		&		&		&		&		 \\ 
c  & 	11100	&	600	&	150	&	300	&	95	&	2	&	50	&	5	&	6.9	&	0.18	&	2.0	&	.16	&	.4	&		 \\ 
d 	 & 	11000	&	1000	&	182	&	436	&	129	&		&	25	&	&	10 	&		&	2.5	&		&		&	1	 \\ 
e 	 & 		&	1300	&	330	&	420	&		&		&		&		&		&		&		&		&		&		 \\ 
f 	 & 	10000	&	3000	&	200	&	730	&	220	&		&	35	&		&	17	&		&		&		&		&		 \\ 
g	 & 	10800	&	2500	&	275	&	700	&	154	&		&		&		&		&		&	18.2	&		&		&	1	 \\ 
h  & 	9120	&		&	331	&	910	&	275	&		&		&		&   148 	&	22.4	&	15.1	&	.14	&	.14	&		 \\ 
\hline
\multicolumn{14}{l}{
\begin{minipage}{12cm}
\footnotesize {

a Bernard Salas et al.  (2001), FIR data from ISO + optical + UV, empirical

b Kwitter \& Henry (1996), optical + UV data, model

c Keyes et al. (1990), optical + UV data, model

d Middlemass (1990), optical + UV + FIR data, model	

e Perinotto et al. (1980),  optical + UV data, empirical 

f P\'{e}quignot et al. (1978), optical + UV + FIR data, model

g Shields (1978),  optical + UV + FIR data, model

h Aller 1954, optical data, empirical
}
 \end{minipage}}\\
\hline
\end{tabular}
\end{flushleft}
\label{Table5}
\caption{Abundances in NGC 7027}
\end{table*}

\begin{table*}
\begin{flushleft}
\begin{tabular}[h]{lrcccccccccccc}
\hline               
 	 & 			He	&	C	&	N	&	O	&	Ne	&	Mg	&	S	&	Cl	&	Ar	&		&		&		&		 \\ 
 	 & 			 	&    	&	 	&	 	&	  	&	  	&	 	&	  	&	  	&		&		&		&		 \\ 
a	 & 		90000	&	219	&	86	&	153	&	9.2	&		&		&		&		&		&		&		&		 \\ 
b	 & 		70000	&	300	&	70	&	180	&	3	&	6.9	&	2.5	&	.1	&	0.5	&		&		&		&		 \\ 
c	 & 	  110000	&		&		&	288	&	52	&		&		&		&	2.7	&		&		&		&		 \\ 
d    & 		72000	&		&	66	&	275	&	13	&		&	3:	&		&	0.8	&		&		&		&		 \\ 
e	 & 		93000	&	616:	&	74	&	436	&	74	&		&	4.2	&	.09	&	2.3	&		&		&		&		 \\ 
f	 & 			&	710	&		&		&		&	25	&		&		&		&		&		&		&		 \\ 
g  & 			&		&	45	&	398	&	19	&		&		&		&		&		&		&		&		 \\ 
h	 & 		$>$ 76000	&	794	&	100	&	760	&	40	&		&		&		&		&		&		&		&		 \\ 
\hline
\multicolumn{14}{l}{
\begin{minipage}{12cm}
\footnotesize {

a Henry et al.   (2000), optical + UV data, model

b Hyung et al. (1994), opt echelle + UV + a few IR data, model

c de Freitas Pacheco et al. (1992), optical data,  empirical

d Gutierrez Moreno (1988), optical data,  empirical	

e Aller \& Czyzak (1983), optical data, hybrid method

f Harrington et al. (1980), optical + UV data, empirical

g Barker (1978), optical data, empirical

h Torres-Peimbert \& Peimbert (1977), optical data, empirical with $t^{2}$ = 
0.035

}
 \end{minipage}}\\
\hline
\end{tabular}
\end{flushleft}
\label{Table6}
\caption{Abundances in IC 418}
\end{table*}

 \subsection{NGC 7027 and IC 418: two test cases}

It is instructive to compare the abundances determined by various 
authors for two bright and well studied PNe. 

NGC 7027 is the PN
with the highest optical surface brightness despite of 
3.5 mag absorption by dust and is a benchmark for PN 
spectroscopists.  It is a very high excitation nebula, with lines of 
[Ne~{\sc vi}] 
now measured (Bernard Salas 2001).  The central star temperature is 
estimated to be 140\,000 --180\,000~K, 
the gas density is around 5~10$^{4}$\cmcub. The nebula is surrounded 
by a dusty neutral shell. Table 5 lists the abundances derived for 
this object. Substantial differences are seen among the 
results obtained by various authors. 
 It is interesting to recall that the 
concommittent models of Shields (1978) and of P\'{e}quignot et al. 
(1978) produced  \oii\ and 
\nii\ intensities about one order of magnitude smaller than 
observed. Multidensity geometries and modifications of the stellar 
continuum failed to resolve this difficulty. P\'{e}quignot et al. 
(1978) postulated the existence of efficient charge transfer 
processes, and obtained an excellent fit to the observations by 
adjusting the charge transfer rates. These charge transfer rates were 
later confirmed by atomic physics computations. In spite of the 
different approaches adopted by Shields ((1978) and P\'{e}quignot et al. 
(1978),  
the resulting abundances are rather similar. On the other 
hand, they are significantly different from the abundances obtained 
later for this object. This is not only due to the use of different
atomic data or to the 
number and quality of observational constraints 
(e.g. ISO spectroscopy provided 
high quality measurements on a large number of IR lines): when models are not entirely 
satisfactory, the abundances finally adopted are a matter of the 
author's personal choice.

IC 418 is also a bright and relatively dense ($n$ $\sim$ 5~10$^{4}$\cmcub) PN,
but with a central star of low effective temperature (\Tstar $\sim$ 38\,000~K), 
so that fewer ions are observed.
The nebula is surrounded by an extended neutral shell. Here again, 
there are substantial differences among the published abundances. In 
this case, the differences in O/H cannot be attributed 
to ionization correction, since O is observed in all its ionization 
stages. It is actually the observational data which strongly differ 
from one author to another! Besides, results from empirical methods 
depend, as we know, on the assumptions made for the temperature 
structure. As for models, they do not give satisfactory fits for 
this object and  therefore do not provide reliable abundances. 

These two examples may serve as a warning that abundances are 
not necessarily as well determined as might be thought from 
error bars quoted in the literature.

 \subsection{What do PN abundances tell us?}

 The chemical composition of PNe envelopes results 
from a mixing of elements produced by the central star and dredged up 
to the surface with the original material out of which the star was 
made. Basically, 
the evolution of the central star can be described as follows (Bl\"{o}cker 
1999,  Lattanzio \& Forestini 1999).  After completion 
of central hydrogen burning through the CNO bicycle, hydrogen burns in 
a shell around the He core.  Due to core contraction the envelope 
expands.  The star evolves towards larger radii and lower effective 
temperatures and ascends the red giant branch (RGB).  During evolution 
on the RGB, the envelope convection moves downward reaching layers 
which have previously experienced H-burning (first dredge up), and 
brings up processed material to the surface.  This material is mainly 
$^{14}$N, $^{13}$C, $^{12}$C, and $^{4}$He (Renzini \& Voli 1981).

The ascent on the giant branch is terminated by ignition of the 
central helium.  The subsequent evolution is characterized by helium 
burning in a convective core and a steadily advancing  
hydrogen shell.  The fusion of helium produces $^{12}$C by the 
triple $\alpha$ process, and this carbon is in turn subject to $\alpha$  capture to form 
$^{16}$O. Eventually the helium supply is totally consumed, leaving a core 
of carbon and oxygen.  The star begins to ascend the giant branch 
again, now called the asymptotic giant branch (AGB).  When a star 
reaches the AGB, it has the following structure:
a CO core, a He burning shell, a He intershell, a H 
burning shell, and a convective envelope.  In stars more massive than 
4\Ms, the envelope penetrates the region where H burning has occured,
dredging up some of its material to the stellar surface (second dredge up). 
During 
this episode, $^{14}$N and $^{4}$He increase, while $^{12}$C and $^{13}$C 
decrease with $^{12}$C/$^{13}$C staying around one, and $^{16}$O slightly
decreases.

While on the AGB, the star experiences further nucleosynthesis.  
Thermal pulses of the He shell induce a flash-driven convection zone, 
which extends from the helium shell almost to the H shell and deposits 
there some $^{12}$C made in the He shell.  As the helium flash dies away, 
the energy 
deposited causes expansion and cooling, and the external convective region 
reaches down the carbon-rich region left after the flash, bringing 
$^{12}$C and $^{4}$He to the star surface (third dredge 
up).
During thermal pulses, elements beyond iron are produced by slow 
neutron capture (s-process).  This requires partial mixing of hydrogen 
into the carbon rich intershell (Lattanzio \& Forestini 1999): these 
protons are captured by $^{12}$C to produce $^{13}$C which later releases 
neutrons via the $^{13}$C($\alpha$,n )$^{16}$O reaction.
For stars above 5\Ms (at solar metallicity)
a second important phenomenon is hot bottom 
burning.  The convective envelope penetrates into the top of the 
H-burning shell.  Temperatures can reach as high as 10$^{8}$~K.  This 
results in the activation of the CN cycle within the envelope, and the 
consequent processing of $^{12}$C into $^{13}$C and $^{14}$N, with the
result that $^{12}$C/$^{16}$O  is smaller than one.

In summary, nucleosynthesis in PNe progenitors mainly 
affects the abundances of He, N and C in the envelope. The He abundance increases 
during the first, second and third dredge up.  The $^{14}$N abundance 
increases during the first, second and third dredge up.  In the case 
of hot bottom burning,  primary N is produced  out of C 
synthesized in the He shell and brought to the H shell 
after the flash.  The $^{12}$C abundance decreases during first and 
second dredge up but increases during third dredge up, and decreases during 
hot bottom burning. From the synthetic evolutionary models of Marigo 
(2001), the resulting enrichment in PNe envelopes with respect to the 
ISM may be as large as a factor of 10 or more for $^{12}$C and $^{14}$N.

The abundance of $^{16}$O is slightly reduced as a consequence of hot 
bottom burning while, as pointed out by Marigo (2001), low mass stars may produce 
positive yields of $^{16}$O, which is brought to the surface by third 
dredge up. Globally, the oxygen abundance is expected to be 
little affected by nucleosynthesis in PN progenitors (Renzini \& 
Voli 1981, 
Forestini \& Charbonnel 1997, van den Hoek \& Groenewegen 1997, Marigo 
2001). From the synthetic evolutionary models of Marigo (2001), the PN 
progenitors modify the PN oxygen abundance by at most a factor of 2, the 
effect being strongest at low metallicities (1/4 solar). At solar and 
half solar metallicity, the effect is practically negligible. 
As a consequence, the abundance of oxygen should be representative of 
the chemical composition of the matter out of which the progenitor 
star was made.  The same holds for the abundances of elements such as 
Ne, Ar, S. On the 
other hand, the abundances of He, C, N and the s-process elements 
tell about the nuclear and mixing processes  in the PN 
progenitors.

When using PNe as indicators of the chemical evolution of 
galaxies, one should be aware that PNe with different 
central star masses probe different epochs and are subject to 
different selection effects. The mere existence of the PN phenomenon 
requires that the star must have reached a temperature sufficient  to 
ionize the surrounding gas before the ejected envelope has vanished 
into the interstellar space. Now, the evolution of the central star is 
more rapid for higher masses. PNe ionized by more massive 
nuclei reach higher luminosities, and they will be the ones for which 
 abundances will be preferentially measured in distant galaxies. In nearby 
 galaxies and in the Milky Way, observations are feasible for 
 lower luminosity PNe. The observability of a PN depends on the 
 detection threshold, but if it is low enough, PNe with less massive 
 nuclei will be visible for a considerably longer time than PNe with 
 massive nuclei. This results from the post-AGB evolution time being a 
 strongly decreasing function of core mass (see e. g. the models of 
 Bl\"{o}cker 1995). Another point is that, because of the existence of an 
 initial-final mass 
 relation (e.g. Weidemann 1987), PNe with less massive nuclei correspond 
 to stars with lower initial masses, which are far more numerous 
 according to the Salpeter initial mass function. For these two reasons,  
 samples of nearby PNe will not contain a large proportion of objects 
 with high mass progenitors. They will not contain many
  PNe with central star masses below 1--1.5\Ms\ either, because such 
 stars are believed to turn into very slowly evolving post-AGB stars 
 and the ejected envelope will have dispersed into the 
 interstellar medium before being ionized. This is why the 
 distribution of central star masses is so strongly peaked around 
 0.6\Ms\ (Stasi\'{n}ska et al. 1997). PNe of different central star 
 masses probe different epochs of galaxy history. Schematically, they 
 can be classified as shown in Table 7 (which however must be taken only 
 as a rough guideline). The subdivision of PNe into four types by 
 Peimbert (1978) was motivated by this kind of considerations 
 (but several revisions to his initial scheme were proposed 
 later, as will be discussed in Sect. 5.4.1). All the above 
 considerations need confirmation from observational data on PNe 
 samples.

\begin{table*}
\begin{flushleft}
\begin{tabular}[h]{rcrr}
\hline               
 		progenitor mass	&	central star mass	&	progenitor's birth	&	PN type$^{a}$		 \\ 
 	 & 			 	&    	&	 		 \\ 
	 2.4 -- 8~M$_{\odot}$    &  $>$ 0.64~M$_{\odot}$  & 1 Gyr  ago    & Type I \\
	 1.2 -- 2.4~M$_{\odot}$  & 0.58 -- 0.64~M$_{\odot}$  & 3 Gyr  ago    & Type II \\
	 1.0 -- 1.2~M$_{\odot}$  &  $\sim$   0.56~M$_{\odot}$  & 6 Gyr ago     & Type III \\
	 0.8 -- 1.0~M$_{\odot}$  &  $\sim$  0.555~M$_{\odot}$  & 10 Gyr ago     & Type IV \\
\hline
\multicolumn{4}{l}{
\begin{minipage}{12cm}
\footnotesize {

${^a}$ PN types according to Peimbert (1978, 1990)

}
\end{minipage}}\\
\hline
\end{tabular}
\end{flushleft}
\label{Table7}
\caption{Schematical classification of PNe and their progenitors}
\end{table*}
  
 \subsection{PNe as probes of the chemical evolution of galaxies }
 
 \subsubsection{The universal Ne/H versus O/H relation}
 
 From a compilation of PNe abundances in the Galaxy and in the 
 Magellanic Clouds, Henry (1989) found that the Ne/H versus O/H 
 relation for PNe is very narrow and linear in logarithm. It is also 
 identical to the one found for \hii\ regions (Vigroux et al. 1987). 
 This implies that Ne and O abundances in intermediate mass stars are 
 not significantly altered by dredge up, and therefore that oxygen and 
 neon abundances in PNe can indeed be used to probe the interstellar 
 abundances of oxygen over a large portion of the history of galaxies. 
  
 \subsubsection{Abundance gradients from PNe in the Milky Way}

\begin{table*}
\begin{flushleft}
\begin{tabular}[h]{lrrrrrrrrrr}
\hline 
	 & 	He	&	C	&	N	&	O	&	Ne	&	S	&	Ar	&	range	&	nb		\\
	 	 & 		&		&		&		&		&		&		&		&	     \\ 
a  & 		&		&	-.084	&	-.054	&	-.069	&	-.064	&		&	5-12	&	43	\\ 
 	 & 		&		&	$\pm$~.034	&	$\pm$~.019	&	$\pm$~.034	&	$\pm$~.035	&		&		&		 \\ 
     & 		&		&		&		&		&		&		&		&		\\  
b	 & 		&		&		&	-.058	&	-.036	&	-.077	&	-.051	&	4-14	&	128	 \\ 
 	 & 		&		&		&	$\pm$~.007	&	$\pm$~.010	&	$\pm$~.011	&	$\pm$~.010	&		&	 \\ 
     & 		&		&		&		&		&		&		&		&		\\  
c 	 & 		&		&		&	-.069	&	-.056	&	-.067	&	-.051	&	4-13	&	91	\\ 
 	 & 		&		&		&	$\pm$~.006	&	$\pm$~.007	&	$\pm$~.006	&	$\pm$~.006	&		&	 \\ 
     & 		&		&		&		&		&		&		&		&		\\  
d 	 & 	-.009	&		&	-.072	&	-.03	&	-.05	&		&		&	1-14	&	277	&	 \\ 
 	 & 	$\pm$~.01	&		&	$\pm$~.024	&	$\pm$~.01	&	$\pm$~.02	&		&		&		&		 \\ 
     & 		&		&		&		&		&		&		&		&		\\  
     & 		&		&		&		&		&		&		&		&		\\  
e  	 & 	-.004	&	.023	&	-.030	&	-.030	&	-.030	&	-.016	&	-.042	&		&	74	\\ 
 	 & 	$\pm$~.003	&	$\pm$~.026	&	$\pm$~.014	&	$\pm$~.010	&	$\pm$~.012	&	$\pm$~.019	&	$\pm$~.013	&		&		 \\ 
     & 		&		&		&		&		&		&		&		&		\\  
f	 & 	-.023	&		&	-.086	&	-.031	&		&	-.082	&	-.072	&	1-11	&	15	\\ 
 	 & 	$\pm$~.0033	&		&	$\pm$~.045	&	$\pm$~.0199	&		&	$\pm$~.027	&	$\pm$~.021	&		&			\\ 
     & 		&		&		&		&		&		&		&		&		\\  
g	 & 	-.011	&		&	-.073	&	-.014	&	.102	&	-.079	&	-.049	&	1-9	&	21	 \\ 
 	 & 	$\pm$~0.003	&		&	$\pm$~.026	&	$\pm$~.016	&	$\pm$~.064	&	$\pm$~.047	&	$\pm$~.021	&		&		 \\ 
     & 		&		&		&		&		&		&		&		&		\\  
h	 & 	-.019	&	-.069	&	-.072	&	.072	&		&	-.098	&		&	7-14	&	42	  \\ 
 	 & 	$\pm$~.003	&	$\pm$~.023	&	$\pm$~.028	&	$\pm$~.012	&		&	$\pm$~.022	&		&		&		 \\ 
\hline
\multicolumn{10}{l}{
\begin{minipage}{12cm}
\footnotesize {
a Martins \& Viegas (2000),	Type II, homogeneous rederivation of 
abundances from compiled intensities   

b Maciel \& Quireza (1999), Type II, abundances compiled from the 
literature

c Maciel \& Koppen (1994), Type II, abundances compiled from the 
literature

d Pasquali \& Perinotto (1993), Type I + II , abundances compiled from 
the literature 

e Amnuel (1993), Type In (according to his classification), 
abondances compiled from the literature

f Samland \& al. (1992), Type II, 	homogeneous observational material 
an automated photoionization model fitting

g K\"{o}ppen \& al. (1991), Type II, homogeneous observational 
material and empirical abundance derivations

h Faundez-Abans \& Maciel (1983), Type II, abundances compiled from 
the literature.	

}
 \end{minipage}}\\
\hline
\end{tabular}
\end{flushleft}
\label{Table8}
\caption{Galactic abundance gradients from PNe d log(X/H) / dR in 
kpc$^{-1}$}
\end{table*}

Table 8 presents a compilation of Galactic abundance gradients 
from PNe in units of d log(X/H) / dR in kpc$^{-1}$. 
Column 9 shows the spanned range of galactocentric  
distances in kpc. Column 10 gives the total number of 
objects used to derive the gradients. Note that, as in the case of 
 \hii\  regions, the  quoted uncertainties in the published abundance 
 gradients include only the scatter in the nominal values of the derived 
abundances. In the case of PNe, distances are not known with good 
accuracy, they are usually derived from statistical methods, 
typically within a 
factor of 2 or more. However, if a gradient is found 
with erroneous distances, this  means that a gradient is indeed most likely 
present, since one does not expect a conspiration of errors in distances 
to produce a spurious gradient. On the other hand, the values of the 
computed gradient strongly depend 
 on the adopted PNe 
distance scale, as noted by Amnuel (1993). Only PNe arising from disk population 
stars are suitable to 
determine abundance gradients in the Galactic disk. Therefore,  high
 velocity PNe (Type III according to the classification by Peimbert 1978)
  and a fortiori PNe belonging to the halo (Type IV PNe) are not 
  suitable for this purpose.

  It is to be noted that, while the existence of gradients seems 
  established, there are   significant differences in the 
 magnitudes of these gradients as  found by different authors.  At 
 present, it is not possible to say how accurate these gradients are. 
 Note that accounting for possible ``temperature fluctuations'' would 
 probably steepen the derived gradients (Martins \& Viegas 2000).
  
From the most recent results,  galactic gradients found for O, Ne and S 
using PNe
appear to be similar to the ones found from \hii\ regions and young 
 stars (Maciel \& Quireza 1999). This suggests that abundance gradients 
 in the Galaxy have not changed during the last 3~Gyr. 
N and C gradients are different between PN and \hii\ regions, which is expected of course. Their 
 values have been reported in Table 8 only to be complete, but the 
 existence of N or C gradients in PNe populations would rather tell 
 something on the stellar populations from which the PN arise. As for 
 the C gradients, they are highly unreliable anyway.
 
We can compare the average O/H in PNe and in \hii\ 
regions of the solar vicinity, using the 
gradients given in Tables 4 and 8 and adopting for simplicity that 
the galactocentric distance of the Sun is 8.5~kpc. We find that 
12+log O/H = 8.81 $\pm$ 0.04 using the \hii\ regions 
data from Shaver et al. (1983),  8.606 $\pm$ 0.06 using those from
Afflerbach et al. (1997), and 8.63 $\pm$ 0.05 using Type II PNe from 
the compilation of  
Maciel \& Quireza (1999).  There is therefore 
no compelling evidence that  O/H differs between Type II PNe 
and \hii\ regions in the solar 
vicinity. This is a further 
indication that ISM abundances have remained constant during the last 
few Gyr and that there is no significant modification of O/H 
 in PNe due to mixing in the progenitors.

 Maciel \& K\"{o}ppen (1994) have examined whether 
 abundance gradients in the Galaxy steepen with time, by comparing the 
 gradients from Type I, Type II and Type III PNe. The evidence is 
 marginal.

The question of possible vertical abundance gradients has been 
investigated by Faundez-Abans \& Maciel (1988), Cuisinier et al. (1996) and 
K\"{o}ppen \& Cuisinier (1997), the latter study being the most 
detailed.  
Adopting careful selection criteria on the quality of the spectra and 
the location of the PNe in the Galaxy in a sample of 94 PNe, 
the latter authors find a systematic decrease of the 
abundances of He, N, O, S and Ar with height above the plane.  The N/O 
ratio also exhibits a clear decrease with height. These findings are 
compatible with a simple empirical model that the authors work out 
for the kinematical and 
chemical evolution of  the solar neighbourhood in which the progenitor 
stars are supposed to be born in  the galactic plane and reach greater heights 
due to the velocity dispersion that increases with age.

 \subsubsection{PNe in the Galactic bulge}

\begin{table*}
\begin{flushleft}
\begin{tabular}{lrrrrrrrrrrrr}
\hline 
ref	 & 		He	&		N	&	O	&	Ne	&	S	&	Ar	&	nb	&		&		&		&		&	 \\ 
	 & 			&	 	&		&		&		&		&		&			&		&		&		&	 \\ 
a	 & 	 	 .101		&	 8.13	&	8.48	&	7.96	&		&		&	85	&		&		&		&		&	 \\ 
	 & 		$\pm$~.028	&	$\pm$~0.42	&	$\pm$~0.43	&	$\pm$~.36	&		&		&		&		&		&		&		&	 \\ 
	 & 			&			&		&		&		&		&		&			&		&		&		&		 \\ 
b	 & 	 	.107			&	8.12	&	8.74	&		&	6.86	& 6.28
		&	30	&		&		&		&		&	 \\ 
	 & 		$\pm$~.019			&	$\pm$~0.37	&	$\pm$~0.15	
	 &		&	$\pm$~0.20	&	$\pm$~0.37	&		& 	&		&		&		&\\		  				

	 & 			&			&		&		&		&		&		&		&		&		&		&		 \\ 
c	 & 	 	.126			&	7.64	&	8.22	&	7.25	&	6.48	
&	5.95	&	45	&	&		&		&		&		 \\ 
	 & 		$\pm$~.027			&	$\pm$~0.55	&	$\pm$~0.43	
	 &	$\pm$~0.46	&	$\pm$~0.93	&	$\pm$~0.55	&		& 	&		&		&		&\\		  				
\hline 
\multicolumn{10}{l}{
\begin{minipage}{12cm}
\footnotesize {
a Stasi\'{n}ska et al. (1998), compiled intensities,
 homogeneous abundance derivations

b Cuisinier et al. (2000), homogeneous data, results kindly 
provided by A. Escudero

c Escudero \& Costa (2001), homogeneous data
}
 \end{minipage}}\\  
 \hline                            
\end{tabular}
\end{flushleft}
\label{Table9}
\caption{Mean abundances of PNe in the Galactic bulge}

\end{table*}

\begin{table*}
\begin{flushleft}
\begin{tabular}[h]{lrrrrrrrrrrr}
\hline

 	 & 	He   	&	N	&	O	&	Ne	&	S	&	Ar	 &		&		&		&   \\ 
	 & 			&		&		&		&		&		&	 &		&		&		&  	  \\ 
a	 & 			& 8.86	&	9.16	&	8.51	&	7.83	&		 &		&		&		&   	 \\ 
	 & 	    	&	$\pm$~0.29	&	$\pm$~0.16	&	$\pm$~0.30	&	$\pm$~0.30	&	 &		&		&		&  	  \\ 
	 & 			&		&		&		&		&		  &		&		&		&  	 \\ 
b	 & 			&		&	9.13	&	8.29	&	7.52	&	6.79	  &		&		&		&  	 \\ 
	 & 			&		&	$\pm$~.05	&	$\pm$~.08	&	$\pm$~.08	&	$\pm$~.08	  &		&		&		&  	\\ 
	 & 			&		&		&		&		&	   &		&		&		&   \\ 
c	 & 			&		&	9.25	&	8.46	&	7.46	&	7.07	 &		&		&		&   	 \\ 
	 & 			&		&	$\pm$~0.05	&	$\pm$~0.06	&	$\pm$~0.05	&	$\pm$~0.07	 &		&		&		&    \\ 

	 & 			&		&		&		&		&	 &		&		&		&  		 	 \\ 
f	 & 	10.91	&		7.74	&	8.73	&		&	6.83	&	6.22  	 \\ 
	 & 	$\pm$~0.014	&		$\pm$~0.22	&	$\pm$~0.09	&		&	$\pm$~0.13	&	$\pm$~0.10	 	 \\ 
	 & 				&		&		&		&		&		  \\ 
g	 & 				&		&	8.81	&		&	6.89	&	6.39	  \\ 
	 & 				&		&	$\pm$~.08	&		&	$\pm$~0.12	&	$\pm$~0.05	 	 \\ 
	 \hline
\multicolumn{8}{l}{
\begin{minipage}{12cm}
\footnotesize {
a Martins \& Viegas (2000) 

b Maciel \& Quireza (1999) 

c Maciel \& K\"{o}ppen (1994) 

f Samland \& al. (1992) 

g K\"{o}ppen \& al. (1991) 

}
 \end{minipage}}\\
\hline
\end{tabular}
\end{flushleft}
\label{Table10}
\caption{Abundances at the galactic center extrapolated from disk PNe}
\end{table*}

 Planetary nebulae offer one the best means to investigate the oxygen 
 abundance in the  Galactic bulge. Table 9 presents the mean oxygen 
 abundances for PNe thought to be physically located in 
 the Galactic bulge, using recent data. In all cases, the abundance 
 derivations were made with \Te -based methods. The abundances 
 derived by  Ratag et al. (1997) have not been included, 
 since many of them rely on modelling of 
 objects with no observed constraint of \Te, and are therefore highly 
suspect (see discussion in Sect. 2.2.1). However, the observations 
from Ratag et al. (1997) were used in the compilation of line 
intensities by Stasi\'{n}ska et al. 
(1998), and the abundances rederived in a consistent way with \Te 
-based methods (for objects for which this was possible). The objects 
of Escudero \& Costa (2001) are newly discovered PNe from the list 
of Beaulieu et al. (1999).
 We see that the abundances of PNe in the Galactic bulge have clearly 
 higher mean values and  dispersions in O/H than the extrapolation 
 from disk PNe towards the Galactic center, which are shown in Table 10. 
 The effect may be even stronger than suggested from these tables, 
 since samples of PNe considered as belonging to the Galactic bulge 
 may actually contain PNe of the disk population that are found 
 physically in the same region as the bulge. 

Combining data on about 100 PNe in 
the Galactic bulge from the works of Cuisinier et al. (2000), 
Webster (1988), Aller \& 
Keyes (1987)  with their own data,  Escudero \& Costa (2001) 
 suggest the existence of a vertical 
abundance gradient in the bulge, with lower O/H at high latitudes.

 \subsubsection{PNe in the Galactic halo}
 
Only a small number of PNe in the halo  are known so far, less than 20, 
for an expected total  of several thousands 
(see e.g.  Tovmassian et al. 2001).  This number is however rapidly 
growing, thanks to systematic sky surveys at high Galactic latitudes for 
the search of emission line galaxies, and in which PNe are discovered 
serendipitously.  Halo PNe belong to an old metal poor 
stellar population, and therefore serve as probes of the halo chemical 
composition at  the time of the formation of their progenitors.  They 
also give the opportunity to study mixing processes in metal poor 
intermediate mass stars.

Using published spectral line data, Howard et al. (1997) rederived the 
chemical composition of 9 halo PNe in a consistent way.  
They found that all had subsolar O/H, the most oxygen poor being K648, 
with log O/H + 12=7.61 (i.e.  about 1/20 of the Anders \& 
Grevesse 1989 solar value). They also found that 
the spread in Ne/O, S/O and Ar/O  is much larger than can be 
accounted for by uncertainties alone.  This scatter in PNe abundances 
is similar to the scatter observed in halo stars 
(Krishnaswamy-Gilroy et al. 1988),
 and suggests that accretion of extragalactic 
material occured during formation of the halo. It must be noted 
however that, among PNe considered to be in the halo, some actually 
probably belong to an old disk population (Torres-Peimbert et al. 
1990).

After the study of Howard et al. (1997), a few other PNe were 
discovered in the halo and their chemical composition analyzed (Jacoby 
et al. 1997, Napiwotzki et al. 1994, Tovmassian et al. 2001).  The most 
spectacular one is SBS 1150+599A (renamed PN G 135.9+55.9), which has 
an  oxygen abundance less than 1/100 
solar (Tovmassian et al. 2001).  This makes it by far the most oxygen poor PN known (and 
perhaps the most oxygen poor \emph {star} known).  One may ask whether the 
oxygen abundance in this object really reflects that of the initial star. 
Indeed, bright giants in metal poor globular clusters seem to present
star to star oxygen abundance variations (see e.g. Ivans et al. 
1999), and mixing processes have been invoked to explain these 
abundance patterns (see 
Charbonnel \& Palacios 2001 for a review). 
One could invoke that a similar process affects the oxygen abundance 
in PN G 135.9+55.9. However, Ne is found to be also 
strongly underabundant in this object (Ne/O $\sim$ 0.3, paper in 
preparation), indicating that this object is indeed extremely 
metal poor. In 
that case, the progenitor must have formed very early in the 
Galaxy but given rise to a PN only recently. 
Alternatively, it could have formed out of infalling metal poor 
material at a more recent epoch.

 \subsubsection{PNe probe the histories of nearby galaxies}

A wealth of data exist for large samples of PNe in the Magellanic 
Clouds, both in the optical and in the UV (Monk et al. 1988, Boroson \& Liebert 
1989, Meatheringham \& Dopita 1991 a, b, Vassiliadis et al. 1992, 
Leisy \& Dennefeld 1996, Vassiliadis et al. 1996, 1998).  PNe in the 
Magellanic Clouds 
represent a statistically significant sample at a common 
distance,  suffering little extinction along the line of sight, and 
 sufficiently bright to allow the measurement of diagnostic lines 
from various ions.

The oxygen abundances of PNe in the Magellanic Clouds
 span a relatively small range: 
log O/H + 12 = 8.10 $\pm$ 0.25 from a compilation of 125 objects for the 
LMC, log O/H + 12 = 7.74 $\pm$ 0.39 from a compilation of 48 objects for 
the SMC reanalyzed in a homogeneous way by Stasi\'{n}ska et al. (1998).  If one 
considers only the high luminosity sample ($L_{{\rm [O~{\sc iii}]}}$ $>$ 100\Ls), the spread 
is smaller and the mean abundance is significantly larger: 8.28 $\pm$ 
0.13 
(40 objects) for the LMC, 8.09 $\pm$ 0.11 (11 objects) for the SMC. This 
has been interpreted as due to the fact that, as a class, high luminosity PNe 
have progenitors of higher masses, therefore younger and made of more 
chemically enriched gas.  The mean oxygen abundance in the high 
luminosity class compares well with that from \hii\ regions in the 
Magellanic Clouds: 
8.35 $\pm$ 0.06 for LMC, 8.03 $\pm$ 0.10 for SMC (Russell \& Dopita 1992).  This 
indicates that the oxygen abundance in luminous PNe is a very good 
proxy of the present day ISM oxygen abundances.  

Dopita et al. (1997) have produced self consistent photoionization 
models to fit the 
observed line fluxes between 1200 and 1800~\AA~for 8 PNe in the LMC. 
With these models they obtain not only the elemental abundances, but also the 
temperatures and luminosities of the central stars. This allows them to 
place the objects in the HR diagram and derive the central star masses 
and post-AGB evolution times by comparison with theoretical tracks 
for post-AGB stars of various masses (the choice of H-burning or 
He-burning track for each object is made by the requirement of 
consistency with the observed expansion age of the nebula).  Assuming the 
initial-final mass relation of Marigo et al. (1996), Dopita et al. 
(1997)  are able to 
estimate the masses of the progenitors. This allows them to 
trace the age-metallicity relationship in the LMC. As a proxy of 
metallicity, they use the sum of the abundances from the 
$\alpha$-process elements Ne, S, Ar (in order to alleviate any doubts 
that might come from the use of O whose abundance can be slightly 
affected by mixing processes). They find that the LMC experienced a 
long period of quiescence, followed by a short period activity within 
the past 3~Gyr which multiplied its metallicity by a factor 2. A 
further study is under way by the same autors to include 20 
additional PNe in the LMC and 10 PNe in the SMC.

PN spectroscopy is now possible with relatively high signal-to-noise 
even in more distant galaxies. 
For example, observations of 28 PNe in the bulge of M31 and 9 PNe in 
the companion dwarf galaxy M32 allowed to obtain \Te-based abundances 
for these objects (Richer et al. 1999). 
The oxygen abundances of the PNe observed in the bulge of M31 are found to be 
very similar to those of the luminous PNe in the Galactic bulge 
(the comparison, in order to be meaningful, must 
be done on nebulae with similar luminosities, since the oxygen 
abundances has been shown to depend on luminosity in the Magellanic 
Clouds and the Galactic bulge). One finds  log O/H + 12 = 8.64 $\pm$ 0.23 
for the M31 bulge sample and 
8.67 $\pm$ 0.21 for the  high luminosity PNe in the Galactic bulge
(Stasi\'{n}ska et al. 1998).
Jacoby \& Ciardullo (1999) obtained spectroscopic data on 12 PNe in the 
bulge and 3 in the disk of M31.  They span a 
larger luminosity range than Richer et al. (1999) who were mainly 
interested in bright PNe. For the three objects in common with Richer 
et al. (1999), the 
oxygen abundances are in excellent agreement.  Yet, for their entire 
sample, Jacoby \& Ciardullo (1999) find log O/H 
+ 12 = 8.50 $\pm$ 0.23 which is significantly lower than the value found by 
Stasi\'{n}ska et al. (1998), possibly because of the larger range of 
PNe luminosities in their sample.

The data on  M32 by Richer et al. (1999)  confirm
 the suggestion by Ford (1983) that the PNe 
in M32 are nitrogen rich. It seems unlikely that all the luminous PNe 
have high enough central star masses to undergo second dredge up, and 
this finding suggests that in M32 nitrogen was already enhanced in 
the precursor stars.

Other local group galaxies have smaller masses and therefore contain 
only a few PNe.  Abundance data exist for PNe in NGC 6822, 
 NGC 205,  NGC 185, Sgr B2, Fornax (see references in Richer \& Mc 
 Call 1995 and Richer et al. 1998).

Richer \& Mc Call (1995) compared the oxygen abundances from PNe in 
diffuse ellipticals and dwarf irregulars.  They found that diffuse 
ellipticals have higher abundances than similarly luminous dwarf 
irregulars. This seems consistent with the idea that diffuse 
ellipticals would be the faded remnants of dwarf ellipticals.  
However, when considering also the O/Fe ratios, obtained by combining 
stellar abundance measurements, they conclude that  diffuse 
ellipticals and dwarf ellipticals have had in fact fundamentally 
different star formation histories.

Combining the data on PNe in these dwarf spheroidals galaxies 
with those on PNe in M32 and in the bulge M31 and of the Milky Way, 
Richer et al. (1998) 
have shown that the mean oxygen abundance correlates very well 
with the mean velocity dispersion. Since the oxygen abundance of 
luminous PNe is a good proxy of the oxygen abundance in the ISM 
at the time when star formation stopped, this implies that there is a 
correlation between the energy input from supernovae and the 
gravitational potential energy.  Such a correlation arises naturally 
if chemical evolution in these systems is stopped by Galactic winds.

The oxygen abundances found in the elliptical 
galaxy NGC 5128 (Centaurus-A) by Walsh et al. (1999) show
 a mean value of about 8.4, i.e. smaller 
than the mean value determined for the bright PNe in M31. This 
result is somewhat difficult to understand for such a massive 
galaxy, unless the most metal rich stars do not produce observable PNe. 
This possibility is known as the 
the AGB manqu\'{e} phenomenon (see e.g. Greggio \& Renzini 1990), by which 
intermediate mass stars do not 
reach the top of the AGB due to intense stellar winds.

 \subsection{PNe probe the nucleosynthesis in their progenitor stars}
 
  \subsubsection{Global abundance ratios}

It is clear from the diagrams presented by Henry et al. (2000) that 
PNe show  significantly higher values of He/H, N/O, C/O than 
\hii\  regions of the same O/H. This indicates that He, N and C have been 
synthesized in PNe progenitors, as theory predicts. More quantitative 
comparison with theory is difficult because of the number of 
determining parameters (stellar mass, parametrization of the mixing 
processes) and  of complex selection effects. 
In the following we draw a few examples of more detailed 
interpretations of abundance ratios that have been proposed.

The nature of Type I PNe is a good example of the difficulty in the 
interpretation.   Peimbert (1978)  had defined type I PNe as objects 
having  He/H $>$ 0.125 and N/O $>$ 0.5. 
Kaler et al. (1978) interpreted the high N/O and He/H 
together with the (He/H , N/O) correlation observed in such objects as due 
to second  dredge up, implying initial stellar masses larger than 
3\Ms.  Later, the He/H criterion to define Type I PNe was abandoned 
(it must be noted that old determinations of  He/H 
did not include proper correction for collisional excitation of He 
lines). Henry (1990) found that Type I PNe showed an (N/O, O/H) anticorrelation
and concluded that in these objects N is produced at the expense of O 
(due to ON cycling).
Kingsburgh \& Barlow (1994) contested the existence of such an anticorrelation 
and propose a new definition of Type I PNe, as being 
PNe that underwent envelope burning conversion to N of dredged up primary 
C. Thus they are objects in which the present N/H is larger than the 
initial (C+N)/H (equal to 0.8 in the solar vicinity).
Costa et al. (2000) on the contrary  define Type I PNe using only the 
criterion He/H $>$ 0.11. They find a (N/O, O/H) anticorrelation when 
PNe are segregated by types. They interpret this by saying 
that the oxygen
 abundance is not modified by the PN progenitor but reflects the 
 metallicity of the site where the progenitor was born, 
 and that dredge-up is more efficient at low metallicity.
 It must be noted that, whatever the definition, there is actually no 
 clear  dichotomy between Type I and other PNe, the distribution of 
 the N/O ratios is rather continuous (and this is also
 what is predicted at least at solar and half solar metallicities from 
 the  models of Marigo 2001). 

Concerning carbon,  (C+N+O)/H is found to increase  with 
C/H and becomes dominated by C/H for the most carbon rich objects.
This is seen both in Galactic samples 
(Kingsburgh \& Barlow 1994) and in 
Magellanic Clouds samples (Leisy \& Dennefeld 1996). This is in
 agreement with a scenario where carbon is produced by 3-$\alpha$
  from He and brought to the surface by third dredge up. From the 
  number of PNe with observed C enhancement, one concludes that 
  third dredge up is common in PNe  progenitors. 
 Among the PNe in which the carbon abundance could be determined, about 40\% 
 (in the Galactic sample) and 70\% (in the Magellanic Clouds sample) 
 have C/O $>$ 1. This is well in line with theoretical predictions that 
 third dredge up is more efficient at low metallicity. Note that PNe 
 with C/O $>$ 1, the so-called carbon-rich PNe, are likely to contain carbon 
 rich dust, since their progenitors must have 
 developed a carbon chemistry to form grains in their atmospheres.

More detailed comparisons of PNe results with the predictions of post 
AGB models have been attempted by Henry et al. (2000) and Marigo (2001). 
Interpretations are difficult, due to the number of parameters 
involved and to the difficulty to derive accurate central star masses 
and to relate them to initial masses.

P\'{e}quignot et al. (2000) discuss two 
PNe in the Sgr B2 galaxy, He 2-436  and Wray 16-423, 
whose nuclei are interpreted as  
belonging to the same evolutionary track.  The authors perform a differential 
analysis of these two PNe, based on tailored photoionization modelling, 
and argue that while systematic errors may substantially shift the derived 
abundances, the conclusions based on \emph{differences} between 
the two models should not be influenced. The main conclusion is that 
third dredge up O enrichment is observed in He 2-436, at the 10 
\%  level. 
  
  \subsubsection{Abundance inhomogeneities}

Many studies have suggested that structures seen in planetary nebulae 
(extended haloes, condensations) have different composition from the 
main nebular body, indicating that they are formed of  material 
arising in distinct mass loss episodes characterized by different 
chemical compositions of the stellar winds. However,  
these  differences in chemical composition may be spurious, due 
inadequacies of the adopted abundance determination scheme. For 
example, the knots and other small scale structures seen in PNe are 
possibly the result of instabilities or magnetic field shaping, and their 
spectroscopic signature could be due to a difference in the excitation 
conditions and not in the chemical composition. 
In the following, some examples of such studies are presented, 
adopting the view of their authors.

a) Extended haloes

NGC 6720, the ``ring nebula'' is surrounded by two haloes: 
 an inner one, with petal-like morphology, and an outer one, 
 perfectly circular, as seen in the pictures of Balick et al. 
 (1992). Guerrero et al. (1997) have studied the chemical 
 composition of these haloes, and found that 
the inner and outer halo seem to have same composition, suggesting a 
common origin:  the red giant wind. On the other hand, 
 the N/O ratio is larger in the main nebula by a factor of
 2, indicating that the main nebula consists of superwind 
 and the haloes of remnants of red giant wind.

NGC 6543, the ``cat eye nebula'' also shows two halo 
structures: an inner one, consisting of perfectly circular rings, and 
an outer one with flocculi attributed to instabilities (Balick et al. 
1992).
Unlike what is advocated for NGC 6720, the rings and the core in NGC 
6543 seem to have same chemical composition 
(Balick et al. 2001). It must be noted however, that the abundances 
may not be reliable, since a photoionization model for the core of NGC 6543
 predicts a far too high \rOiii\ (Hyung et al. 2000). Another puzzle 
 is the information provided by $Chandra$. Chu et al. (2001) estimated that
 the abundances in the X-ray emitting gas are similar 
 to those of the fast stellar wind and larger than 
 the nebular ones. On the other hand, the  temperature of the X-ray gas 
 ($\sim$ 1.7~10$^{6}$~K) is  lower by two orders of magnitudes 
 than the expected post shock 
 temperature of the fast stellar wind. This would suggest that the 
 X-ray emitting gas is dominated by nebular material. These findings 
 are however based on a crude analysis and more detailed model 
 fitting is necessary.

b) FLIERs and other microstructures 

A large number of studies have been devoted to microstructures in 
PNe, and their nature is still debated.
Fast Low Ionization Emission Regions (FLIERs) have first been 
considered to show an enhancement of N and were 
interpreted as being recently expelled from the star 
 (Balick et al. 1994). However, 
Alexander \& Balick (1997)   realized 
that the use of traditional ionization correction factors may lead to 
specious abundances. Dopita (1997) made the point that enhancement of \Nii/\Ha\ 
can be produced by shock compression and does not necessarily involve  
an increase of the nitrogen abundance. 
Gon\c{c}alves et al. (2001) have summarized data on the 50 PNe known to have low 
ionization structures (which they call LIS) and presented a detailed 
comparison of model predictions with the observational 
properties. 
They conclude that not all cases can be satisfactorily explained by 
existing models.

c) Cometary knots

The famous cometary knots  of the Helix nebula NGC 7293 have been 
recently studied by O'Dell et al. (2000) using spectra and images 
obtained with the HST. The 
\Nii/\Ha\ and \Oiii/\Ha\ ratios were shown to decrease with distance to the 
star. Two possible interpretations were offered. Either this could be 
the consequence of a larger electron temperature close to the star due 
to harder radiation field. Or the
 knots close to the star would be more metal-rich, in which case they
 could be relics of blobs ejected during  the AGB stage rather than formed during 
 PN evolution. Obviously,  a more thorough discussion is needed, 
 including a detailed modelling to reproduce the 
 observations before any conclusion can be drawn.

d) Planetary nebulae with Wolf-Rayet central stars 

About 8\% of PNe possess a central star having Wolf-Rayet charateristics, 
with H-poor and C-rich atmospheres. The evolutionary status of these 
objects is still in question. A late helium flash giving rise to a 
 ``born-again'' 
 planetary nebula, following a scenario proposed by Iben et al. 
 (1983), can explain only a small fraction of them. 
 The majority  appear to form an evolutionary sequence from late to 
 early Wolf-Rayet types, 
 starting  from the AGB (G\'{o}rny \& Tylenda 
2000, Pe\~{n} at et al. 2001). This seemed in contradiction with 
theory which predicted that departure from the AGB during a late 
thermal pulse does not produce H-deficient stars. 
Recently however, it has been shown that convective overshooting
can produce a very efficient dredge up, and models including this 
process are now able to produce H-deficient post-ABG stars 
following a thermal pulse on the AGB (Herwig 2000, 2001, 
see also  Bl\"{o}cker et al. 
2001).
It still remains to explain why late type Wolf-Rayet central stars 
seem to have atmospheres richer in carbon than early type ones 
(Leuenhagen \& Hamann 1998, Koesterke 2001).
 Also, one would expect the 
chemical composition of PNe with Wolf-Rayet central stars to be different 
from that of the rest of PNe. This does not seem to be the case, 
as found by G\'{o}rny \& Stasi\'{n}ska (1995), basing on a 
compilation of published abundances: PNe with Wolf-Rayet central stars
are indistinguishable from other PNe in all respects except for
 their larger expansion velocities.  Pe\~{n}a et al. (2001) obtained a 
 homogeneous set of high spectral resolution optical spectra 
 of about 30 PNe with 
 Wolf-Rayet central stars and reached a similar conclusion, as far as 
 He and N abundances are concerned. Their data did not allow to draw 
 any conclusion as regards the C abundances. 

 e) H-poor PNe
 
 There are only a few  PNe  which show the presence of 
 material processed in the stellar interior. They 
 are referred to as H-poor PNe, although the H-poor material is 
 actually embedded in an H-rich tenuous envelope.
 The two best known cases are A 30 and A 78, whose knots are 
 bright in \Oiii\ and \Heii\ but in which Jacoby (1979) could not 
 detect the presence of H Balmer lines. With deeper spectra, 
 Jacoby \& Ford (1983) estimated the 
 He/H ratio to be $\sim$ 8 in these two objects.
Harrington \& Feibelman (1984) obtained IUE spectra of a knot in A 30, and found 
that the high C/He abundance implied by {C~{\sc ii} $\lambda$4267 
is not apparent in the UV spectra, 
suggesting that the knot contains a cool C-rich core. 
Guerrero \& Manchado (1996) obtained spectra of the diffuse nebular 
body of A30, showing it to be H-rich. A similar conclusion was 
obtained 
by Manchado et al. (1988) and  Medina \& Pe\~{n}a (1999) for the outer 
shell of A 78. 
However,  quantitatively, the results obtained by these two sets of 
authors 
are quite different and a deeper analysis is called for. 

Three other objects  belong to this group: A 58, IRAS1514-5258 
and IRAS 18333-2357, the PN in the globular cluster M22, already 
mentioned in Sect. 3.7.5.

One common characteristic of this class of objects is their extremely 
high dust to gas ratio, and the fact that the photoelectric effect on 
the grains provides an important (and sometimes 
dominant) contribution to the heating of the 
nebular gas (see Harrington 1996). This may lead to large 
point-to-point temperature variations (see Sect. 3.7.5) and strongly affect
abundance determinations.

Harrington (1996) concludes his review on H-poor PNe by noting that the 
H-poor ejecta cannot be explained by merely taking material with typical 
nebular abundances and converting all H to He. There is additional 
enrichment of C, N, perhaps O, and most interestingly, of Ne.
 However, more work on 
these objects is needed -- and under way (e.g. Harrington et al. 1997) 
-- before the abundances can be considered reliable. 
Stellar atmosphere analysis of H-deficient central stars (e.g. Werner 
2001) is providing complementary clues to the nature and evolution 
of these objects.

In conclusion, we have shown how nebulae can provide powerful tools to 
investigate
the  evolution of stars and to probe the chemical evolution of galaxies.
Nevertheless, is necessary to
keep in mind the uncertainties and biases involved in the process of 
nebular abundance derivation. These are not always easy to make out, 
especially for the non specialist.
 One of the aims of this review was to help in maintaining a critical eye on the 
 numerous and outstanding achievements of nebular Astronomy.

  \begin{acknowledgments}
	  
ACKNOWLEDGMENTS: It is a pleasure to thank the organizers of the XIII 
Canary Islands 
Winterschool, and especially C\'{e}sar Esteban, for having given me 
the opportunity to share my experience on abundance determinations in 
nebulae. I also wish to thank the participants, for their attention 
and friendship.
I am grateful to Miriam Pe\~{n}a, Luc 
Jamet and Yuri Izotov for a detailed reading of this manuscript, to Daniel Schaerer for 
having provided
useful information on stellar atmospheres and to Andr\'{e} Escudero for 
having kindly computed a few quantities related to planetary nebulae in the 
galactic bulge. I would like to thank my collaborators and friends, 
especially Rosa Gonz\'{a}lez Delgado, Slawomir G\'{o}rny, Claus Leitherer, 
Miriam Pe\~{n}a, 
Michael Richer, Daniel Schaerer, Laerte Sodr\'{e}, Ryszard Szczerba 
and 
Romuald Tylenda for numerous  and lively discussions in various parts of the 
World.

Finally, I would like acknowledge the possibility of
a systematic 
use of the NASA ADS Astronomy Abstract Service
 during the preparation of these lectures.
\end{acknowledgments}

\end{document}